\documentclass[amsmath,amssymb,showpacs,showkeywords,twocolumn]{revtex4}
\usepackage{graphicx}
\usepackage{times}
\usepackage{mathrsfs}
\usepackage{amsmath}
\usepackage{graphicx}
\usepackage{epsfig}
\usepackage{dcolumn}
\usepackage{bm}
\usepackage{multirow}
\usepackage{pdfpages}
\usepackage[utf8x]{inputenc}
\DeclareUnicodeCharacter{2212}{-}
\begin{document}
\title{Field free Josephson diode effect in Ising Superconductor/Altermagnet Josephson junction}
\author{Arindam Boruah\footnote{arindamboruah@dibru.ac.in}, Saumen Acharjee\footnote{saumenacharjee@dibru.ac.in} and Prasanta Kumar Saikia\footnote{saikiapk@dibru.ac.in}}
\affiliation{Department of Physics, Dibrugarh University, Dibrugarh 786 004, 
Assam, India}

\begin{abstract}
Altermagnets (AMs) are an exotic class of antiferromagnet that exhibit spin-splitting even at the absence of net global magnetization and spin-orbit coupling (SOC) effects. In this work, we investigated theoretically, the supercurrent nonreciprocity in an Ising Superconductor/Altermagnet/Ising Superconductor (ISC/AM/ISC) Josephson junction which revealed asymmetric Josephson critical currents, $0 - \pi$ transitions and anomalous current-phase relationship (CPR). A strong Josephson diode efficiency (JDE) is observed due to the combined effects of AM strength and orientations in a conventional SC even in absence of SOC. However, it significantly enhances in presence of intrinsic SOC (ISOC), resulting in pronounced diode effect in both single and double band ISC/AM based Josephson junction. Additionally, it is observed that JDE is more prominent at higher AM strengths with intermediate orientations in all scenario. Notably, it is significantly suppressed for orientations $0^\circ$ and $45^\circ$. Our results also indicate that barrier transparency and AM lengths play a crucial role in optimizing the JDE. In a single-band ISC/AM system JDE persists for any AM length, while reduces at longer AM junction in case of a double-band ISC/AM system. Moreover, our results suggest that a diode efficiency of $\sim 52\%$ can be achieved in the proposed Josephson junction in both single and double band ISC/AM Josephson junction by considering strong AM strength. Furthermore, single band ISC offers wide AM orientation range in contrast to double band ISC for better tunability and optimization of JDE. Our findings highlight the impact of AM strength, orientation and ISOC on the JDE efficiency offering insights for superconducting diode design. 
\end{abstract}

\maketitle

\section{Introduction}
The interplay between superconductivity and magnetism in hybrid Josephson junctions has been a subject of extensive research due to its fundamental significance and potential applications in superconducting spintronics and quantum computing \cite{bergeret1,buzdin1,bergeret2,linder101,saxena,aoki,pfleiderer, zutic,acharjee101}. Although ferromagnet-superconductor junctions
have been extensively studied, revealing exotic transport phenomena such as spin-polarized Andreev reflection \cite{andreev}, long-range
triplet pairing \cite{meng,trifunovic,annunziata}, and the formation of $\phi$-junctions \cite{acharjee1021,acharjee1022,assouline1,yuan1,mayer}, the inherently
detrimental interaction between spin-singlet superconductivity and ferromagnetic ordering has remained a major challenge \cite{hirai}.  
Recently, a new class of magnetic materials, known as Altermagnet (AM) has been  identified in compounds such a RuO$_2$, MnRe, Mn$_5$Si$_3$, MnF$_2$ and La$_2$CuO$_4$ \cite{feng1,occhialini,betancourt,smejkal1,yuan2,smejkal2,moreno}, offering novel possibilities for unconventional transport. AMs breaks time-reversal symmetry ($t \nrightarrow -t$) in a fashion similar to ferromagnet while following alternating magnetic ordering akin to antiferromagnet, resulting in vanishing macroscopic net magnetization \cite{smejkal1,smejkal2,yuan2}. Unlike conventional ferromagnets and antiferromagnets, where spin splitting is dictated by exchange interactions, AMs exhibit a momentum-dependent spin-split band structure governed by their crystal symmetries, reminiscent of the d-wave superconducting order parameter. Furthermore, the absence of macroscopic net magnetization suppresses stray magnetic fields, which are typically problematic in magnetic heterostructures, thereby making AMs promising candidates for next-generation superconducting spintronics applications \cite{cheng11,sun1,ouassou11,papaj,beenakker1}.

One of the most intriguing manifestations of superconducting transport asymmetry  is the Josephson Diode Effect (JDE), wherein the Josephson current exhibits nonreciprocity ($J(\phi) \nrightarrow -J(-\phi)) $, allowing dissipationless supercurrent flow in one direction while being suppressed in the opposite direction \cite{ando,lin,narita,ilic,he,daido1,daido2,legg,zinkl,hou,picoli,hosur}. This effect is of great technological significance for superconducting logic circuits, nonreciprocal quantum devices, and energy-efficient superconducting electronics 
\cite{nadeem,volkov1,zhang11}. Various mechanisms have been proposed to achieve JDE, including Rashba spin-orbit coupling (RSOC) superconductors \cite{ilic3,debnath,matsuo}, asymmetric spin-momentum locking states \cite{fu1}, magnetic superlattices \cite{ando33} and honeycomb lattice based superconductor junction \cite{wei} . However, the realization of a robust JDE in altermagnetic Josephson junctions remains an open question. Given that Ising Superconductors (ISCs), such as two-dimensional monolayer transition-metal dichalcogenides (TMD) (e.g., NbSe$_2$, MoS$_2$, TaS$_2$, WSe$_2$) lacks inversion symmetry, host strong intrinsic spin orbit coupling (ISOC) that protects Cooper pairs against in-plane magnetic perturbations \cite{acharjee1021,barrera,idzuchi3,jalouli2,tang2,lu2,saito2,xi2,dvir2,costanzo2, sohn2, li2, hamil2, ai2}.  
Moreover, the absence of inversion symmetry ($\mathbf{r} \nrightarrow -\mathbf{r}$) and strong SOC, results in an effective Zeeman field at the $(K ,-K)$ valleys that exceeds the Pauli
limit, facilitating long-range superconductivity \cite{lu2, saito2, xi2, dvir2, costanzo2, sohn2, li2, hamil2, ai2,zhu22, xiao2,kormanyos,cappelluti}. These unique properties suggest that the interplay between ISCs and an altermagnet could yield a highly tunable
and unconventional diode effect.

The investigation of the Josephson diode effect (JDE) in ISC/AM/ISC junctions is of considerable interest because of its potential to unveil novel superconducting transport phenomena arising from the interplay between Ising superconductivity and altermagnetism. This system provides a unique platform to explore spin-dependent Andreev bound states (ABS) and their impact on superconducting transport, offering insights beyond conventional spin-singlet or spin-triplet pairing mechanisms \cite{sun1,ouassou11,papaj,beenakker1}. Unlike previously reported JDEs that rely on Rashba spin-orbit coupling or external magnetic fields \cite{ilic3,debnath,matsuo}, the ISC/AM/ISC junction exhibits a diode effect intrinsically governed by the AM's spin texture, leading to a robust and highly tunable Josephson rectification effect. The ability to control this JDE by tuning key junction parameters, such as the altermagnetic strength, interface transparency, and orientation, makes this system highly promising for energy-efficient superconducting logic devices and quantum information applications \cite{nadeem,volkov1,zhang11}. Furthermore, the spin-polarized nature of the the altermagnetic bands suggests the emergence of unconventional ABS, giving rise to novel spin-dependent transport characteristics that could be exploited in superconducting spintronics.  Moreover, the realization of an ISC/AM/ISC
junction provides a unique alternative mechanism that does
not rely on external magnetic fields or strong relativistic effects. Instead, asymmetric superconducting transport is
intrinsically driven by momentum-dependent spin splitting in the altermagnet \cite{papaj,sun1}, which modifies ABS
and the current-phase relation (CPR).

The paper is structured as follows: In Sec. II, we define Bogoliubov–de Gennes Hamiltonian and propose theoretical framework for the ISC/AM/ISC Josephson junction. In Sec. III, we analyze the current-phase relation of the proposed junction and investigate the resulting diode efficiency. In Sec. IV, we discuss our findings and their implications. Finally, Sec. V presents our conclusions.

\section{Formulation}
\begin{figure}[hbt]
\centerline
\centerline{ 
\includegraphics[scale=0.42]{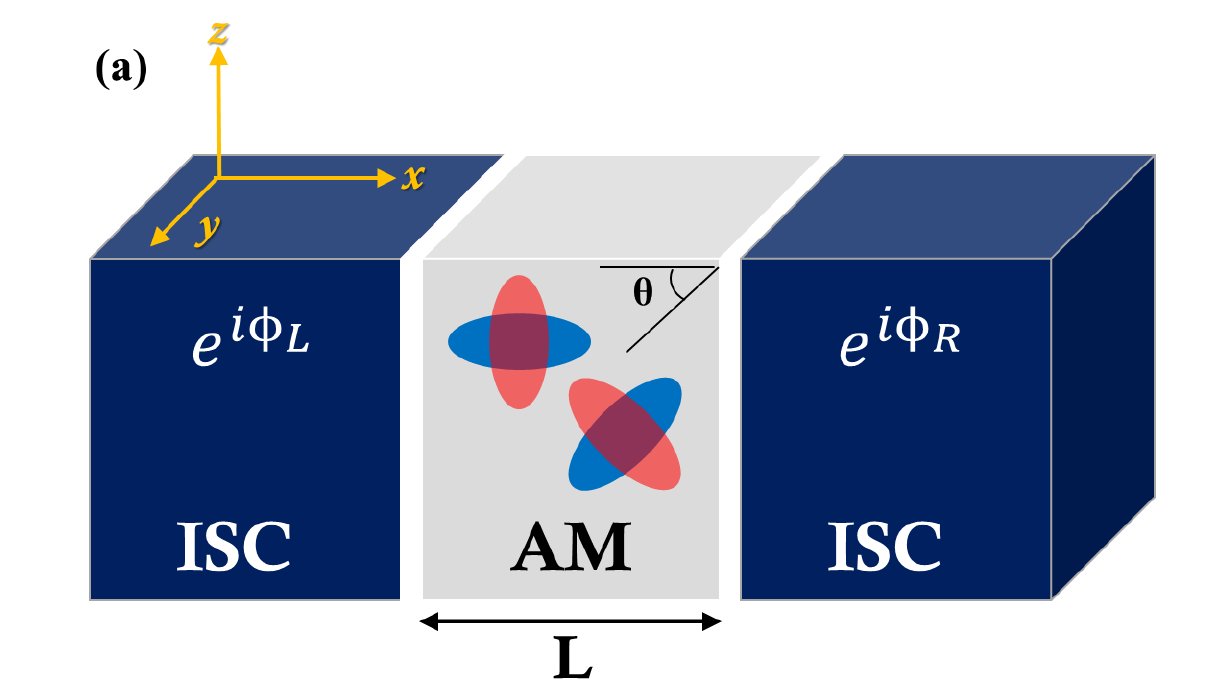}
\vspace{-1.cm}
\includegraphics[scale=0.65]{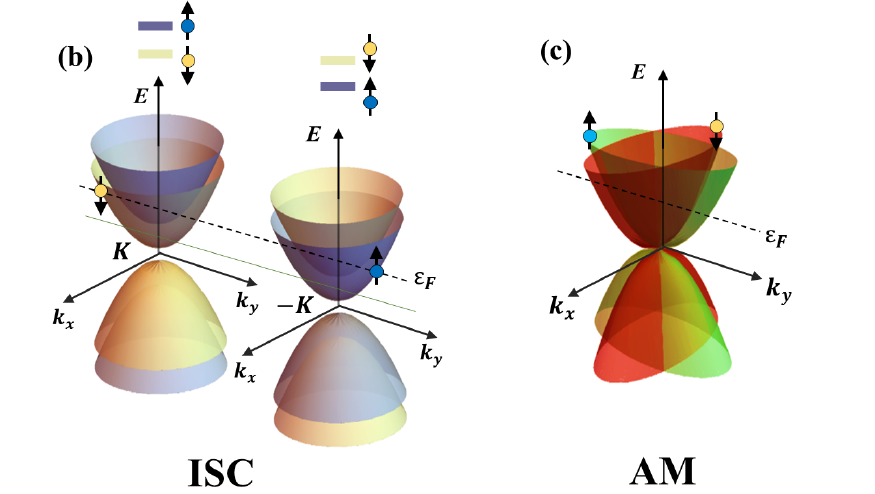}
\vspace{0.6cm}
}
\caption{(a) Schematic illustration of a Josephson junction consisting of a pair of Ising Superconductors (ISC) in close proximity with an Altermagnet (AM). Here, $\theta$ represents different interface orientations, effectively rotating the spin-resolved Fermi surfaces in the AM for the majority (blue ellipse) and minority (red ellipse) spin carriers. (b) Schematic energy band structure of an ISC near the $\pm K$ valleys. The blue (yellow) solid bands indicate the up (down) spin configuration of a double-band (left) and single-band (right) ISC.  The fermi energy $\varepsilon_\text{F}$ of double and single band ISC are denoted by the black dotted and green solid line respectively. (c) Schematic energy band structure of altermagnet. The black dashed lines corresponding to Fermi energy of the AM. }
\label{fig1}
\end{figure}

We consider a Josephson junction geometry consisting of a pair of Ising Superconductors (ISC) ($x < 0$ and $x > \text{L})$ connected by an arbitrary angle oriented Altermagnet (AM) ($0 < x < \text{L})$ as shown in Fig. \ref{fig1}(a).  The AM junction has length $L$ and width $W$. For a sufficiently large width $W$ and in the absence of impurity scattering, translational invariance along the $y$-direction can be assumed, allowing the transverse momentum $k_y$ to be treated as a good quantum number. In this work, we present our analysis for different length schemes considering different values of $L/L_0$, where $L_0$ is considered to be the superconducting coherence length. 

Using a field operator basis, $\Psi = (\psi_\uparrow, \psi_\downarrow, \psi_\uparrow^\dagger, \psi_\downarrow^\dagger)^\text{T}$, the excitation spectrum can be described by the Bogoliubov de Gennes (BdG) Hamiltonian $\check{\mathcal{H}}_\text{BdG} = \check{\mathcal{H}}_\text{ISC} + \check{\mathcal{H}}_\text{AM}$. For the ISC region, the Hamiltonian is defined as  \cite{acharjee1021, cheng11} 
\begin{equation}
\label{eq1}
\check{\mathcal{H}}_\text{ISC} (\mathbf{k}) = \left(
\begin{array}{cc}
 \hat{\mathcal{H}}_\pm(\mathbf{k})   & \hat{\Delta}(\mathbf{k}) \\
 -\hat{\Delta }^{\ast}(\mathbf{k}) & -\hat{\mathcal{H}}^{\ast}_\pm(\mathbf{-k})
 \\
\end{array}\right)
\end{equation} 
\noindent where $\hat{\mathcal{H}}_\pm(\mathbf{k})$ is the single particle Hamiltonian which in the presence of the two valleys ($\mathbf{K}$ and $-\mathbf{K}$) of ISC can be expressed as \cite{tang2}

\begin{equation}
\label{eq2}
\hat{\mathcal{H}}_\pm(\mathbf{k}) = \left(-\frac{\hbar^2\nabla^2}{2m}-\mu_\text{S}\right)\hat{\text{I}} +\varepsilon\beta\hat{\sigma}_z 
\end{equation}
\noindent where $\mathbf{k}$ are the wave vectors of the electron corresponding to the valley $\pm \mathbf{K}$, $\mu_\text{S} = \mu_\text{S}^\text{L}\Theta(-x) + \mu_\text{S}^\text{R}\Theta(x-\text{L})$ is the chemical potentials of the ISC and $\varepsilon = \pm $ is the valley index for $\pm \mathbf{K}$ valleys. The parameter $\beta$ characterize the strength of ISOC and $\hat{\sigma}_z$ is the z - component of Pauli spin matrix. The energy dispersion of the ISC is shown in Fig. \ref{fig1}(b). The up and down spin subbands split into constituents as a result of the ISOC.  It is to be noted that the spin-up band has higher energy than spin-down bands at $\mathbf{K}$ valley, while an opposite characteristic is observed for $-\mathbf{K}$ valley. The normal phase of the ISC still preserves time-reversal symmetry and spin-rotation symmetry along the z-axis. It is to be noted that as the $\mathbf{K}$ and $-\mathbf{K}$ valleys are situated at the edges of the Brillouin zone and are significantly separated. So, the intervalley scattering caused by impurities remains weak and thus can be neglected .
For $\mu_\text{S} > \beta$, the ISC is a double-band superconductor while for $\mu_\text{S} < \beta$, it can behave like a single-band one \cite{acharjee1021}. The Cooper pairs are formed by electrons with the opposite spin and opposite wave vectors from different valleys Fig. \ref{fig1}(b).

\begin{figure*}[hbt]
\centerline
\centerline{
\includegraphics[scale=0.2]{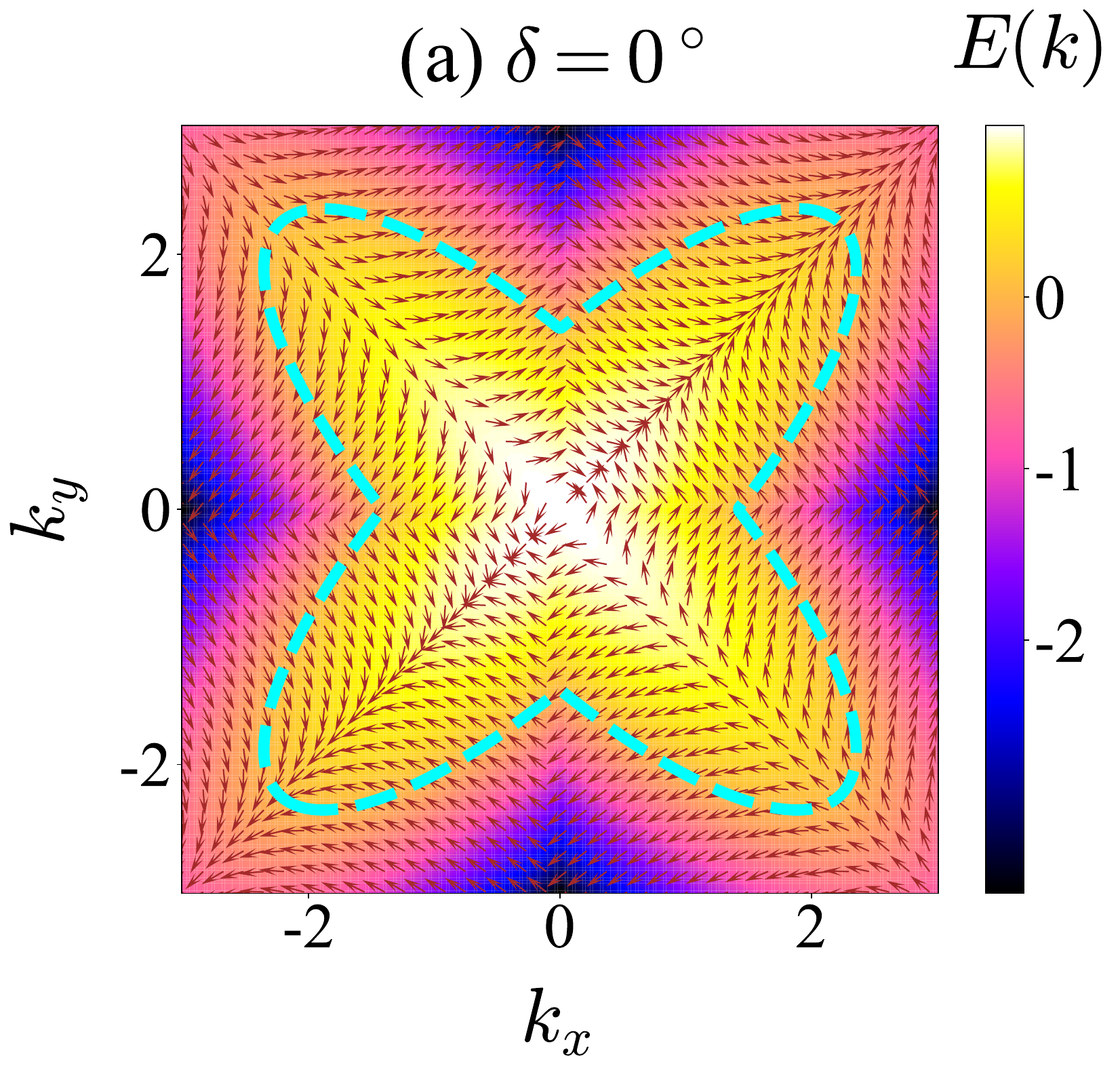}
\vspace{0.01cm}
\includegraphics[scale=0.2]{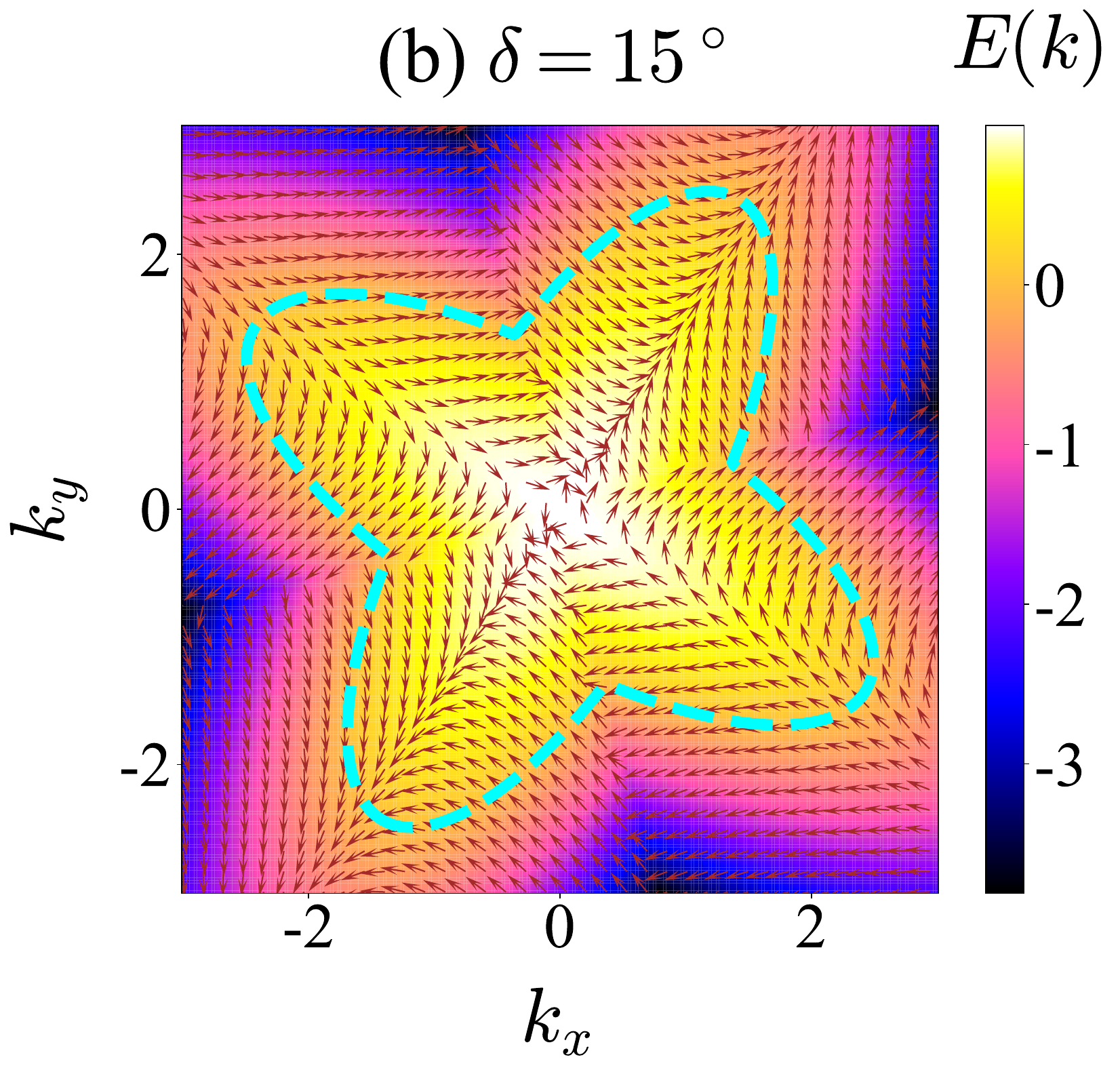}
\vspace{0.01cm}
\includegraphics[scale=0.2]{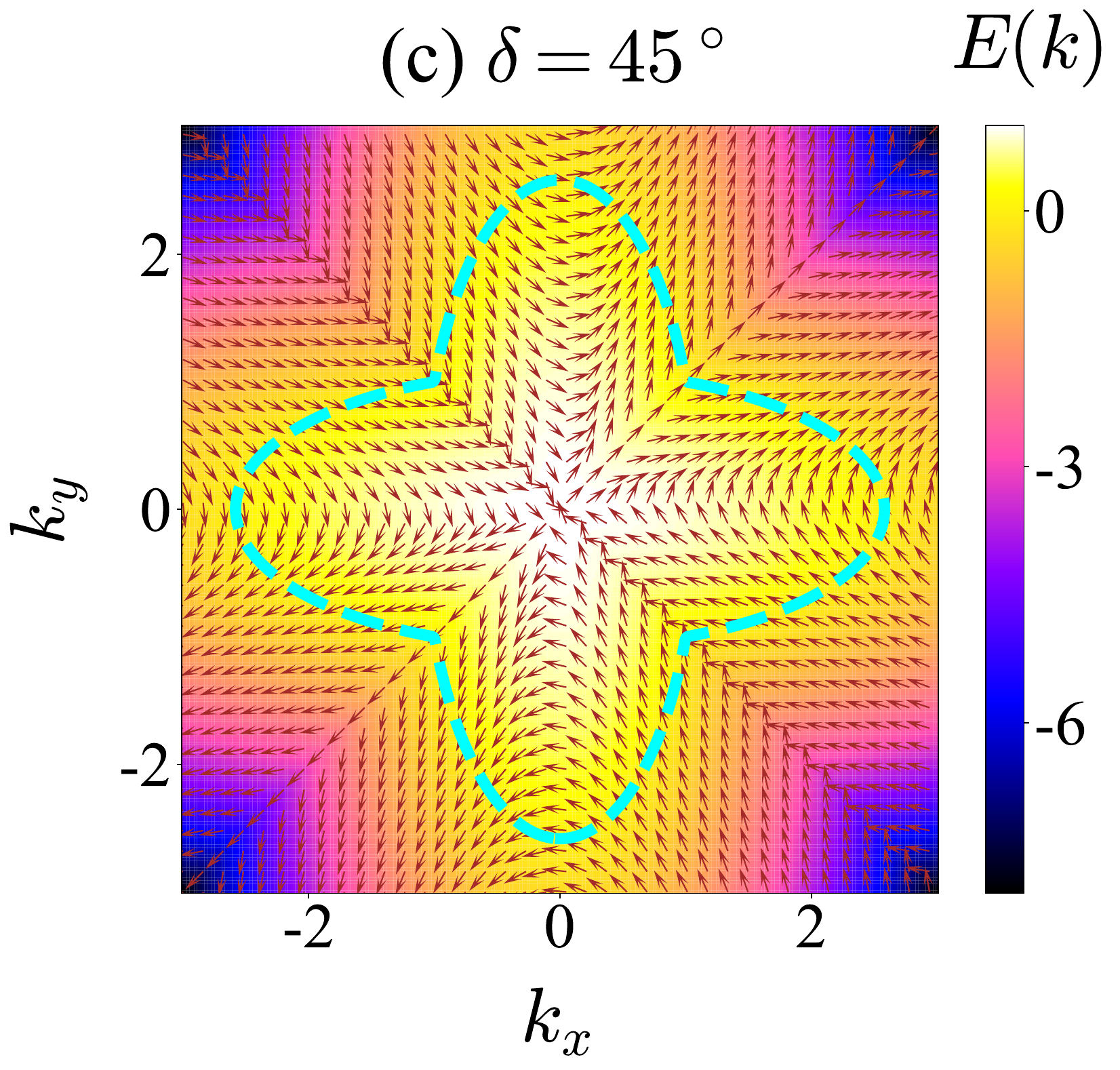}
\vspace{0.01cm}}
\caption{Fermi surfaces and spin structures of electrons in AM for (a) $\delta = 0^\circ$, (b) $\delta = 15^\circ$ and (c) $\delta = 45^\circ$. The Fermi surfaces are denoted by cyan dotted line while brown arrows indicate the spins of the particle.}
\label{fig2}
\end{figure*} 

The superconducting order parameter for the ISC appears in Eq. (\ref{eq1}) can be defined as \cite{tang2,cheng11}
\begin{equation}
\label{eq3}
\hat{\Delta}(\mathbf{k}) = \Delta i \hat{\sigma}_y\lbrace e^{i\phi_\text{L}} \Theta(-x) + e^{i\phi_\text{R}} \Theta(x-\text{L})\rbrace,
\end{equation}
\noindent where $\Delta$ is the magnitude of the superconducting gap and $\phi_\text{L} (\phi_\text{R})$ corresponds to the superconducting phase of the left (right) ISC. Thus, the phase difference between the respective superconductor is $\phi = \phi_\text{L} - \phi_\text{R}$.

The Hamiltonian for an arbitrary angle oriented AM can be written as \cite{papaj}
\begin{equation}
\label{eq4}
\check{\mathcal{H}}_\text{AM} (\mathbf{k}) = \left(
\begin{array}{cc}
\hat{\mathcal{H}}_0(\mathbf{k})   & 0 \\
 0 & -\sigma_y\hat{\mathcal{H}}^\ast_0(\mathbf{-k})\sigma_y 
 \\
\end{array}\right)
\end{equation} 
\noindent where, $\hat{\mathcal{H}}_0(\mathbf{k})$ is the single particle AM Hamiltonian, can be expressed as 
\begin{equation}
\label{eq5}
\hat{\mathcal{H}}_0(\mathbf{k}) = \left(-\frac{\hbar^2\nabla^2}{2m}-\mu_\text{AM}\right)\hat{\text{I}} + \alpha_1 k_xk_y \hat{\sigma}_z + \frac{\alpha_2}{2} (k_x^2 - k_y^2)\hat{\sigma}_z 
\end{equation}
\noindent where, $\mu_\text{AM}$ is the chemical potential in the AM region. $\alpha_1$ and $\alpha_2$ are the dimensionless parameters characterize the d-wave altermagnetic exchange interaction and depends on angle $\delta$ of the AM/ISC interface relative to the crystalline axes. The AM angle is defined as, $\delta = \frac{1}{2}\tan^{-1}(\frac{\alpha_1}{\alpha_2})$. It is to be noted that the AM parameters should satisfy the condition $\bar{\alpha} \equiv \sqrt{\alpha_1^2+\alpha_2^2} < \alpha_c \equiv \frac{\hbar^2}{m}$ to ensure that the Fermi surface is elliptic rather than hyperbolic. he corresponding energy band structure of the AM is depicted in Fig. \ref{fig1}(c). Fig. \ref{fig2} illustrates the elliptic Fermi surface with spin-dependent orientations in momentum space. When the AM is aligned at $\delta = 0^\circ$, spin-polarized transport along the $x$-direction is absent, as shown in Fig. \ref{fig2}(a). However, for $\delta = 45^\circ$, the major axis of the spin-up Fermi surface lobe aligns with the $x$-axis [Fig. \ref{fig2}(c)], thereby can host spin-polarized current flow along the $x$-direction. More generally, for an arbitrary AM orientation $\delta$, the system sustains a finite spin current, demonstrating its capability to support spin-polarized transport, as illustrated in Fig. \ref{fig2}(b).

The wave function in the ISC regions can be obtained by solving the BdG equation using plane wave assumption and diagonalizing the Hamiltonian from Eq.(\ref{eq1}), the wave function in the left ISC can be defined as \cite{acharjee1021, cheng11}

\begin{multline}
\label{eq6}
\Psi^{\text{L}\pm}_{\text{ISC}}(x<0) = 
a^\pm_1\left(u e^{i\phi_{\text{L}}/2}\hat{\chi}_1 + v e^{-i\phi_{\text{L}}/2}\hat{\chi}_4\right)e^{-ik_\pm x}\\
+a^\pm_2\left(u e^{i\phi_{\text{L}}/2}\hat{\chi}_2 - v e^{-i\phi_{\text{L}}/2}\hat{\chi}_3\right)e^{-ik_\mp x}\\+
a^\pm_3\left(v e^{i\phi_{\text{L}}/2}\hat{\chi}_1 
+ u e^{-i\phi_{\text{L}}/2}\hat{\chi}_4\right)e^{ik_\pm x}\\
+a^\pm_4\left(-v e^{i\phi_{\text{L}}/2}\hat{\chi}_2 
+u e^{-i\phi_{\text{L}}/2}\hat{\chi}_3\right)e^{ik_\mp x}
\end{multline}
Similarly, the wave function of the right ISC can also be defined as
\begin{multline}
\label{eq7}
\Psi^{\text{R}\pm}_{\text{ISC}}(x>\text{L}) = 
b^\pm_1\left(u e^{i\phi_{\text{R}}/2}\hat{\chi}_1 + v e^{-i\phi_{\text{R}}/2}\hat{\chi}_4\right)e^{ik_\pm x}\\
+b^\pm_2\left(u e^{i\phi_{\text{R}}/2}\hat{\chi}_2 - v e^{-i\phi_{\text{R}}/2}\hat{\chi}_3\right)e^{ik_\mp x}\\+
b^\pm_3\left(v e^{i\phi_{\text{R}}/2}\hat{\chi}_1 
+ u e^{-i\phi_{\text{R}}/2}\hat{\chi}_4\right)e^{-ik_\pm x}\\
+b^\pm_4\left(-v e^{i\phi_{\text{R}}/2}\hat{\chi}_2 
+u e^{-i\phi_{\text{R}}/2}\hat{\chi}_3\right)e^{-ik_\mp x}
\end{multline}
\noindent where we define, $\hat{\chi}_1 = (1,0,0,0)^\text{T}$, $\hat{\chi}_2 = (0,1,0,0)^\text{T}$, 
$\hat{\chi}_3 = (0,0,1,0)^\text{T}$ and
 $\hat{\chi}_4 = (0,0,0,1)^\text{T}$. Here,  $a^\pm_1$ ($a^\pm_2$) and  $a^\pm_3$ ($a^\pm_4$) represent the reflection 
 coefficients for up (down) spin electrons and holes respectively while  $b^\pm_1$ ($b^\pm_2$)  and $b^\pm_3$ ($b^\pm_4$) are the transmission coefficients for up (down) spin electrons and holes respectively.  The quasi particle momenta of the electron and hole in the ISC can be obtained by diagonalizing Eq. (\ref{eq1}), So it can be expressed as
 \begin{equation}
 \label{eq8}
 k_{e(h),\pm} = \sqrt{\frac{2m}{\hbar^2}(\mu_\text{S} \mp \epsilon \beta +\tau \sqrt{E^2-\Delta^2)}}
\end{equation}   
where, $\tau = + (-)$ for electron (hole) considered for our analysis. Here, $k_{+(-)}$ represents the momenta of the electron (hole) in the ISC. Under Andreev approximation, we can neglect the contribution of $\sqrt{E^2-\Delta^2}$. which yields an error of the order $\frac{\delta k_\text{FS}}{k_\text{FS}} = \frac{\sqrt{E^2-\Delta^2}}{\mu_\text{S}}$ which is of the order of $\frac{\Delta}{E}$. However, as $\Delta \ll E$, so it can be considered as a valid approximation \cite{acharjee1021,cheng11}.  The momenta of the electron and the holes under the Andreev approximation can be written as $
k_{e(h),\pm} \sim \sqrt{\frac{2m}{\hbar^2}\left(\mu_\text{S} \mp \beta\right)}$

The quasiparticle amplitudes $u$ and $v$ appearing in Eq.(\ref{eq8}) are defined as
\begin{equation}
\label{eq9}
 u =  \frac{1}{\sqrt{2}}\sqrt{1+\sqrt{1-\frac{\Delta^2}{E^2}}}; \,\,\,\,\,\,\,
 v =  \frac{1}{\sqrt{2}}\sqrt{1-\sqrt{1-\frac{\Delta^2}{E^2}}}
\end{equation}
The wave function in the AM region can be defined as 
\begin{multline}
\label{eq10}
 \Psi_{\text{AM}\pm}(0 < x < L) = c^\pm_1\hat{\chi}_1e^{i\kappa^+_{+}x} + c^\pm_2\hat{\chi}_2e^{i\kappa^+_{-}x} 
\\ + c^\pm_3\hat{\chi}_3e^{-i\kappa^-_{+}x} + c^\pm_4\hat{\chi}_4e^{-i\kappa^-_{-}x}
\end{multline}
\noindent where $c^\pm_1$, $c^\pm_2$, $c^\pm_3$, $c^\pm_4$ are the scattering coefficients for electron and hole in the AM region with $\pm$ corresponding to the wave function $ \Psi_{\text{AM}\pm}$ respectively. The quasi-particle momenta in this region are given by 
\begin{equation}
\label{eq11}
\kappa^\pm_{e(h)} =\frac{-m\alpha _1 \kappa_y\pm\sqrt{2 m \mathcal{P}_1(\mu_\text{AM}+E )  + \kappa_y^2 \mathcal{P}_2}}{\mathcal{P}_1}
\end{equation}
\noindent where we define, $\mathcal{P}_1\equiv  \left(\hbar^2+\tau m\alpha _2 \right)$ and $\mathcal{P}_2 \equiv  \left(\alpha _1^2+\alpha _2^2\right) m^2-\hbar^4$. The parameter,  $\mu_\text{AM}$ represent the chemical potential of the AM. An extensive form of the AM wavevectors and the density of states of the AM region is presented in Appendix (\ref{appenA}) and (\ref{AppenB}) respectively.

The boundary conditions can be derived by applying $\check{\mathcal{H}}_\text{BdG}\Psi = E\Psi$ and integrating over the interval $[-\epsilon, \epsilon]$, in the limit $\epsilon \rightarrow 0$. Furthermore, the hermiticity of the Hamiltonian operator is preserved by enforcing the antisymmetrization of the altermagnetic term. 
\begin{figure*}[hbt]
\centerline
\centerline{
\includegraphics[scale = 0.475]{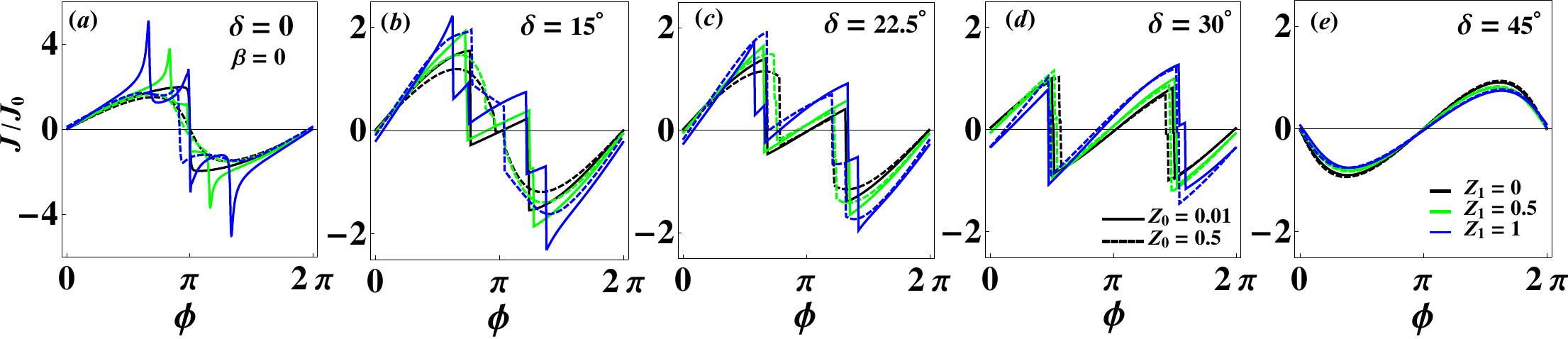}}
\caption{Josephson Supercurrent ($J/J_0$)  as a function of phase difference ($\phi$) for different values of $\delta$ in absence of ISOC ($\beta = 0$) considering the AM strength $Z_1 = 0$ (black), $Z_1 = 0.5$ (green) and $Z_1 = 1$ (blue) line with $\mu_S = 1$ and $L/L_0 = 1$.  Solid lines correspond to a barrier strength of $Z_0 = 0.01$, while dashed lines represent $Z_0 = 0.5$.}
\label{fig3}
\end{figure*} 
Consequently, the wave functions are required to satisfy the following boundary conditions:
\begin{align}
\label{eq12}
\Psi^{\text{L}}_{\text{ISC}\pm}(0^-) &= \Psi_{\text{AM}\pm}(0^+)   ,\\
\label{eq13}
\Psi_{\text{AM}\pm}(\text{L}^-) &= \Psi^{\text{R}}_{\text{ISC}\pm}(\text{L}^+),\\
\label{eq14}
\hat{\Lambda}_1\partial_x\Psi_{\text{AM}\pm}(0^+)-\partial_x\Psi^{\text{L}}_{\text{ISC}\pm}(0^-)
&= Z_0\Psi^{\text{L}}_{\text{ISC}\pm}(0^-),\\
\label{eq15}
\partial_x\Psi^{\text{R}}_{\text{ISC}\pm}(\text{L}^+)-\partial_x\Psi_{\text{AM}\pm}(\text{L}^-)
&= \hat{\Lambda}_2\Psi_{\text{AM}\pm}(\text{L}^-)
\end{align}
\noindent where we define the vectors as, $\hat{\Lambda}_1 = [(1+\Gamma_2), (1-\Gamma_2), (1-\Gamma_2), (1+\Gamma_2)]^\text{T}$, $ \hat{\Lambda}_2 = [(Z_0+i Z_1), (Z_0-i Z_1),(Z_0-i Z_1),(Z_0+i Z_1)]^\text{T}$ with $Z_1 = \Gamma_1-\Gamma_2 \partial_x$ considering $\Gamma_1 \equiv \frac{i\alpha_1 k_y}{\hbar^2}$ and $\Gamma_2 \equiv \frac{\alpha_2 m}{\hbar^2}$. The magnitude of the  parameters $Z_0 = \frac{2mU_0}{\hbar^2}$ and $|Z_1| = \sqrt{\Gamma_1^2 + k_x^2 \Gamma_2^2}$ characterize the barrier strength and AM strength respectively. It is to be noted that both $Z_0$ and $Z_1$ are expressed as dimensionless quantities normalized by the Fermi energy $E_F$.

\section{Andreev levels and Josephson Supercurrent} 
The Andreev levels in the ISC$\vert$AM$\vert$ISC Josephson junction can be determined by imposing the boundary conditions given in Eqs. (\ref{eq13})–(\ref{eq16}). This leads to a system of equations of the form $\hat{\mathcal{M}}\hat{x} = 0$, where $\hat{\mathcal{M}}$ is a $16 \times 16$ matrix encapsulating all the information about the bound state energies, and the vector $\hat{x} = (a^\pm_1, a^\pm_2, a^\pm_3, a^\pm_4, c^\pm_1, c^\pm_2, c^\pm_3, c^\pm_4, c^{\pm'}_1, c^{\pm'}_2, c^{\pm'}_3, c^{\pm'}_4, b^\pm_1, b^\pm_2, b^\pm_3,$ $b^\pm_4)^\text{T}$ represents the set of coefficients associated with the wave functions. The bound state energies and the corresponding Andreev levels are obtained by solving the condition $\det[\hat{\mathcal{M}}(E_\pm)] = 0$, where $\det[\dots]$ denotes the determinant of the matrix. A detailed expression for $\hat{\mathcal{M}}(E_\pm)$ is provided in Appendix (\ref{appenC}).  
\begin{figure*}[hbt]
\centerline
\centerline{
\includegraphics[scale = 0.477]{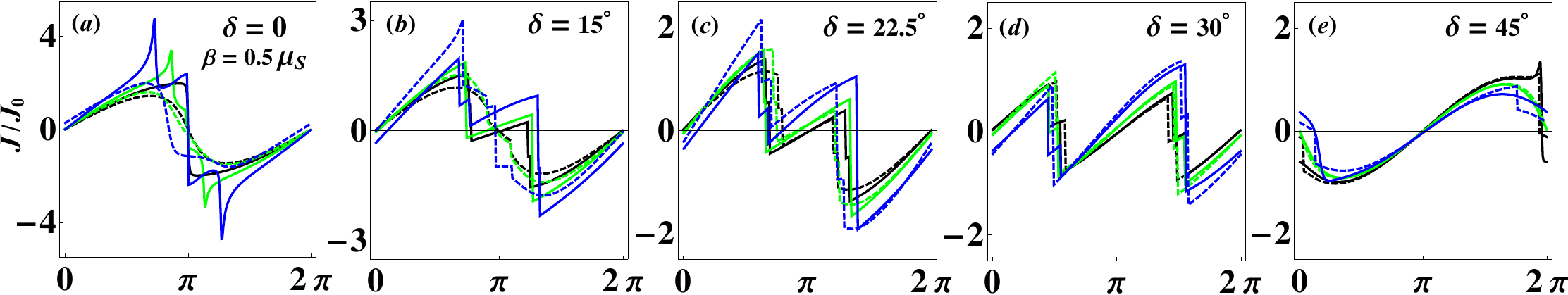}
\includegraphics[scale = 0.4745]{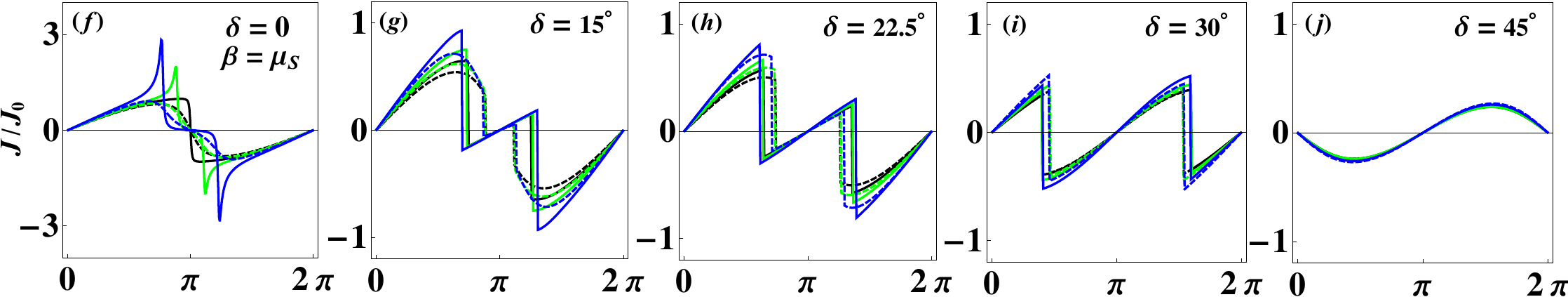}
\includegraphics[scale = 0.477]{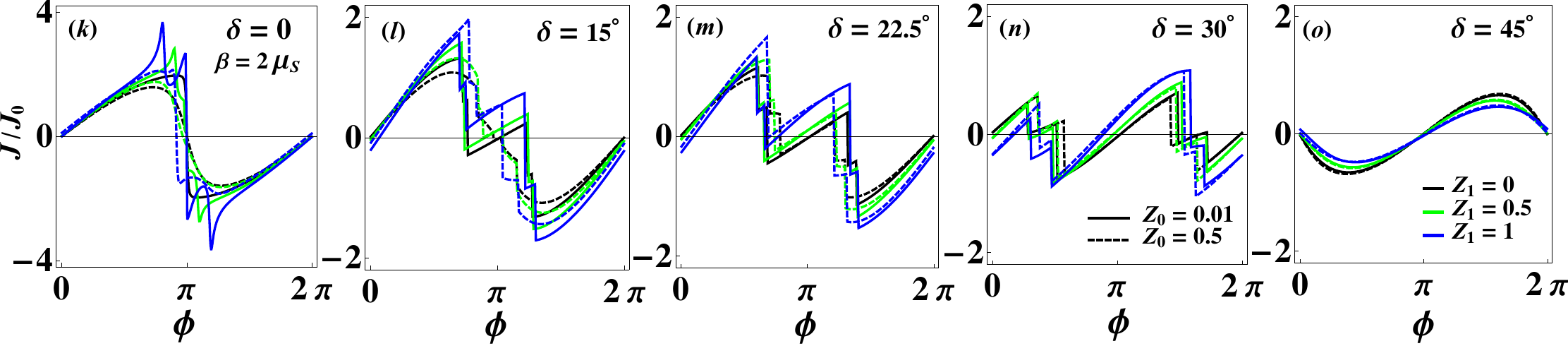}}
\caption{Josephson Supercurrent ($J/J_0$) as a function of phase difference ($\phi$) for different values of $\delta$ in presence of ISOC considering the AM strength $Z_1 = 0$ (black), $Z_1 = 0.5$ (green) and $Z_1 = 1$ (blue).  Solid lines correspond to a barrier strength of $Z_0 = 0.01$, while dashed lines represent $Z_0 = 0.5$. The CPR are shown for a single-band ISC with $\beta = 0.5 \mu_S$ (top panel), intermediate $\beta = \mu_S$ (middle panel), and for a double-band ISC with $\beta = 2 \mu_S$ (bottom panel).
}
\label{fig4}
\end{figure*} 
An expression for the normalized Josephson supercurrent can be obtained by using the standard expression \cite{cheng11,acharjee1021}
\begin{equation}
\label{eq16}
J(\phi,\delta) = J_0\int_{-\frac{\pi}{2}}^{\frac{\pi}{2}} \cos\theta d\theta\sum_\sigma \tanh\left(\frac{ E_{\sigma}}{2k_BT}\right) \frac{dE_\sigma}{d\phi}
\end{equation}  
where we define, $J_0 = 2e\Delta/\hbar$ and $E_\sigma$ corresponds to the energies of the Andreev levels with $\sigma = \pm 1$. Here, $\theta$ is the angle of incidence of the incoming electron. We set, $T = 0.001T_c$  with $T_c$ represents the critical temperature of the ISC and $\mu_S = 1$. However, to understand the impact of $\mu_S$ we also examine the current-phase relation (CPR) for different values of $\mu_S$.  The Josephson current has a $k_F$ dependence for large values of $Z_0$. To capture different barrier configurations, we consider a transparent junction with $Z_0 = 0.01$ and an opaque junction with $Z_0 = 0.5$. Furthermore, we investigate the impact of AM strength in CPR and JDE considering the parameter $Z_1$ in three distinct regimes: weak ($Z_1 = 0.1$), moderate ($Z_1 = 0.5$), and strong ($Z_1 = 1$).

\subsection{Current Phase Relation in absence of ISOC}
Before we present our CPR results for ISC/AM/ISC junctions, it is essential to first understand the impact of AM on a conventional SC/AM/SC Josephson junction in the absence of SOC. Fig. \ref{fig3} illustrates the behavior of the supercurrent in an AM-based Josephson junction characterized by different AM strength $Z_1$ and orientation $\delta$, considering two barrier transparencies, $Z_0 = 0.01$ and $Z_0 = 0.5$, with $\beta = 0$ and $L/L_0 = 1$.  For $\delta = 0$ with $Z_1 = 0$, the system exhibits the characteristics of a conventional Josephson junction for both transparent and opaque barrier regime, as seen in Fig. \ref{fig2}(a). However, for a nearly transparent barrier ($Z_0 = 0.01$) and moderate AM strength ($Z_1 = 0.5$), the CPR displays pronounced anharmonicity, characterized by sharp peaks and discontinuities. This behavior suggests the presence of unconventional Andreev reflections and the emergence of higher harmonics in the system \cite{golubov, blu, olund}. As the AM strength further increases to $Z_1 = 1$, the CPR exhibits even stronger anharmonicity and nonlinearities in the Andreev bound states (ABS), indicating nontrivial phase shifts and a highly nonlinear supercurrent response solely due to AM.  
For an opaque barrier ($Z_0 = 0.5$), the CPR regains a predominantly sinusoidal behavior even in the presence of moderate AM strength ($Z_1 = 0.5$). This characteristics arises from the suppression of higher-energy ABS due to the increased tunneling resistance. However, for a strong AM strength ($Z_1 = 1$), the CPR again deviates from normal sinusoidal form, indicating a resurgence of nontrivial supercurrent contributions. Notably, in the absence of AM ($Z_1 = 0$), the system satisfies $J(\phi) = -J(-\phi)$, confirming the absence of  JDE.  Fig. \ref{fig3}(b) presents the CPR for $\delta = 15^\circ$. Notably, the system exhibits a $\phi$-phase Josephson junction for this orientation when $Z_0 = 0.01$. However, distinct $\phi$ phases are observed for $Z_1 = 0$ and $Z_1 = 1$, indicating that the $\phi$-state can be tuned via the AM strength. In this regime, the CPR features sharper and more asymmetric peaks, accompanied by a significant phase shift, suggesting the emergence of a JDE. This behavior can be attributed to the arbitrary orientation of the Fermi surface lobes, as illustrated in Fig. \ref{fig2}(b). Although the sharpness of the peaks are significantly reduced for $Z_0 = 0.5$ with $Z_1 = 0$ and $0.5$, but a deviation from non sinusoidal behaviour is noticed for $Z_1  = 1$. Similar characteristics with more promising asymmetric behaviour in CPR  are observed in Fig. \ref{fig3} (c) and Fig. \ref{fig3}(d), as $\delta$ increases to $22.5^\circ$ and $30^\circ$. However, the sharp transitions gradually weaken \cite{blu, olund}. This trend suggests that a larger orientation angle suppresses higher-order harmonics, leading to smoother CPR curves \cite{golubov}. Furthermore, the JDE remains robust for both $Z_0 = 0.01$ and $Z_0 = 0.5$, highlighting its potential tunability. Further rise in $\delta  = 45^\circ$, results in a $\pi$-phase Josephson junction as indicated by Fig.\ref{fig3} (e). Moreover, the system display more conventional sinusoidal CPR with suppressed diode effect and reduced contributions from higher harmonics for both $Z_0 = 0.01$ and $0.5$, which is due to dominance of the majority spin AM band. Thus, asymmetric CPR and consequent JDE can be achieved even in conventional SC/AM junction with arbitrary orientation and strong AM strength.

\subsection{Current Phase Relation in presence of ISOC}
In presence of ISOC, the wave vectors  $k_+$, and $k_-$ of ISC are significantly different leading to strong spin splitting of the Andreev energy bands.  Fig. \ref{fig4}(a)-\ref{fig4}(e) present the CPR for single band ($\beta = 0.5 \mu_S$)  while the CPR for double band ($\beta = 2\mu_S$) ISC are shown in Fig. \ref{fig4}(k)-\ref{fig4}(o) for different values of $Z_0$, $Z_1$ and $\delta$. Although $J(\phi)$ shows almost identical pattern in this condition as observed previously for $\beta = 0$ from Fig. \ref{fig3}, the presence of ISOC breaks the inversion symmetry, along with its interplay with $Z_1$ and $\delta$ modifies the ABS spectrum, leading to higher asymmetry in positive and negative supercurrents in the junction. As observed from Fig. \ref{fig4}(a) and Fig. \ref{fig4}(k), for small $\delta$, the CPR exhibits strong non-sinusoidal behavior with higher harmonics, thereby enhancing the diode efficiency. The asymmetric nature of supercurrent are even more pronounced for $\delta = 15^\circ$, $22.5^\circ$ and $30^\circ$ for both $\beta < \mu_S$ and $\beta > \mu_S$ as seen from Figs. \ref{fig4}(b) - \ref{fig4}(d)  and Figs. \ref{fig4}(l) - \ref{fig4}(n).  The impact of ISOC can be visualized from Fig. \ref{fig4}(d) and Fig. \ref{fig4}(n), where, a non-reciprocity in supercurrents is noticed even at $Z_1 = 0$, which was absent for $\beta = 0$. It signifies that ISOC together with an arbitrary AM orientation can enhance the asymmetry in $J(\phi)$. It is observed that the effect is more prominent in the single-band regime due to stronger spin-momentum interaction, leading to an enhanced forward-backward asymmetry in the critical currents $J_c^+$ and $J_c^-$ \cite{cheng11}. 
\begin{figure}[hbt]
\centerline
\centerline{
\includegraphics[scale = 0.52]{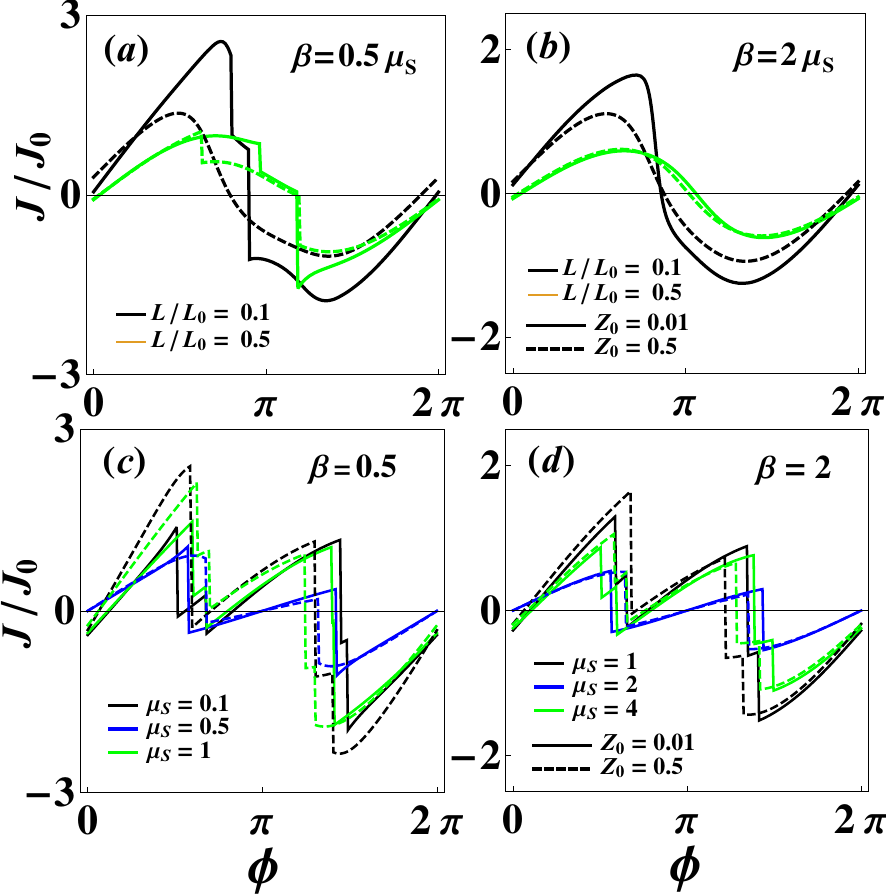}}
\caption{Top panel: $J/J_0$ as a function of $\phi$ for (a) single band ISC ($\beta = 0.5\mu_S$) and (b) double band ISC ($\beta = 2\mu_S$) with $L/L_0 = 0.1$ (black) and $L/L_0 = 0.5$ (green) considering $Z_1 = 1$, $\delta = 22.5^{\circ}$ and $\mu_S = 1$. The solid and dashed lines are for $Z_0 = 0.01$ and $0.5$ respectively.
Bottom panel: $J/J_0$ as a function of $\phi$ for (c) single band ISC ($\beta = 0.5\mu_S$) with $\mu_S = 0.1$ (black), $\mu_S = 0.5$ (blue) and $\mu_S = 1$ (green) and (d) double band ISC ($\beta = 2\mu_S$) with $\mu_S = 1$ (black), $\mu_S = 2$ (blue) and $\mu_S = 4$ (green) considering  $Z_1 = 1$, $\delta = 22.5^{\circ}$ and $L/L_0 = 1$.}
\label{fig5}
\end{figure}

In contrast, in the double-band regime, the significant Fermi surface mismatch suppresses the interband coupling, yet the diode effect persists due to momentum-selective transmission across the AM barrier. It is due to the interplay of spin-dependent phase shifts, spin filtering at the interface, and the asymmetric modification of the superconducting phase gradient due to finite ISOC-induced spin splitting. However, for $\delta = 45^\circ$, the CPR approaches to sinusoidal form as seen from Fig. \ref{fig4}(e) and Fig. \ref{fig4}(o), indicating reduced diode efficiency, which can be attributed due to destructive interference between spin-split ABS branches. The dependence on $Z_1$ further highlights the role of interface transparency in modifying the diode effect, with lower transparency enhancing asymmetry due to suppressed interband scattering. Furthermore, for $\beta = 2\mu_S$, the system undergoes a $0-\pi$ transition of the junction for all $\delta$ orientations while for $\beta = 0.5\mu_S$ still follows $\phi$-phase configuration. In the intermediate region ($\beta = \mu_S$), the sharp peaks and the $0-\pi$ transition of the junction is seen as well. Moreover, the critical currents $J_c^+$ and $J_c^-$ are found to be highly symmetrical, exhibiting no JDE as seen from Figs. \ref{fig4}(f)-\ref{fig4}(j). The Fermi surfaces of the spin-split ISC bands become symmetric with respect to the superconducting gap, leading to an equal contribution of spin-up and spin-down quasi-particles to the supercurrent. Consequently, the spin-dependent phase shifts, which are responsible for asymmetric CPR and nonreciprocal supercurrent cancel out, resulting in a purely sinusoidal CPR. Additionally, the higher-order harmonics necessary for nonreciprocal transport are suppressed due to the cancellation of spin-dependent phase shifts, and thus restores inversion symmetry in the CPR. This behavior underscores the fundamental role of Fermi surface asymmetry, ISOC and AM strength interplay in determining the diode efficiency in unconventional Josephson junctions.
\begin{figure*}[hbt]
\centerline
\centerline{
\includegraphics[scale = 0.2]{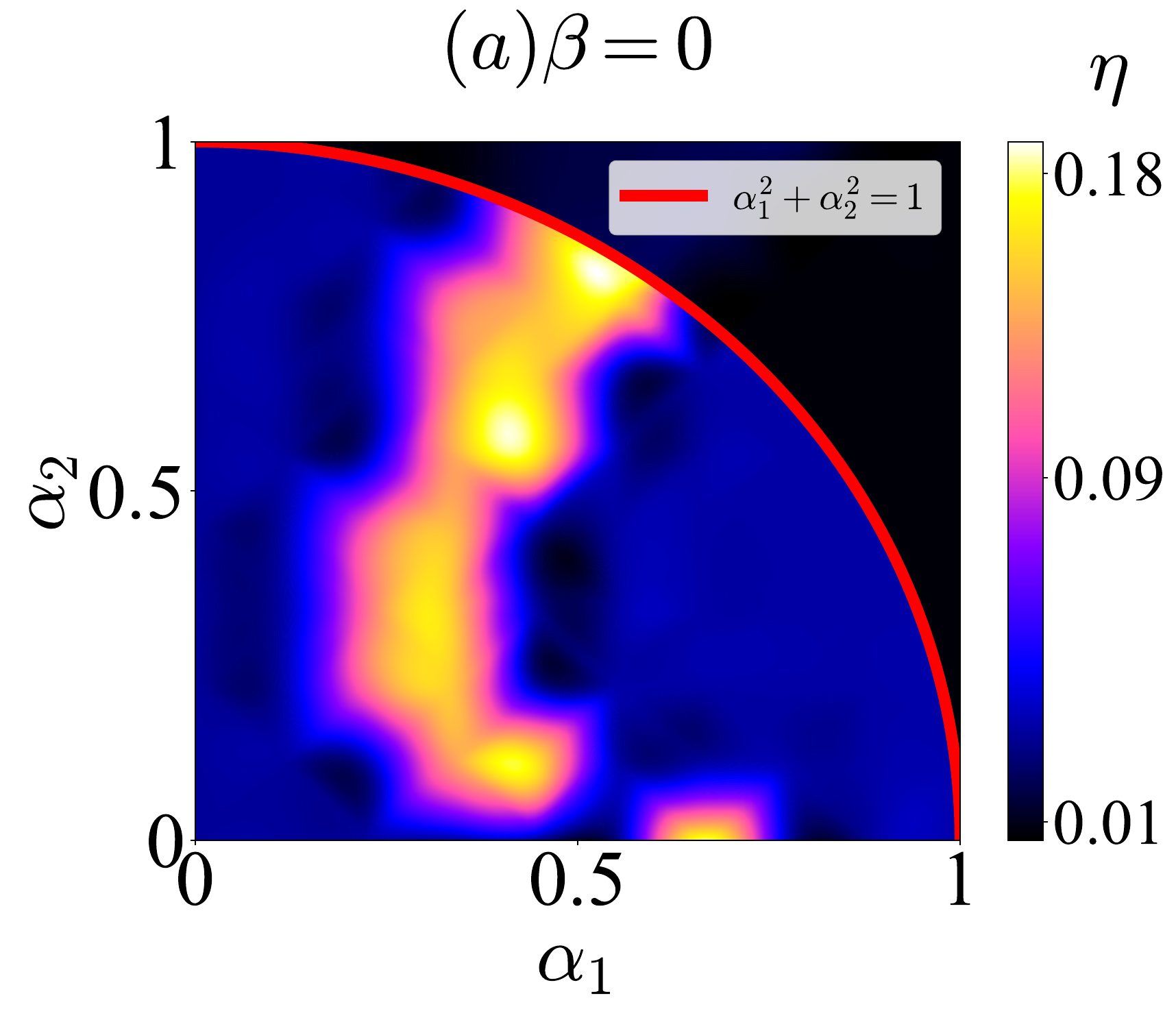}
\includegraphics[scale = 0.2]{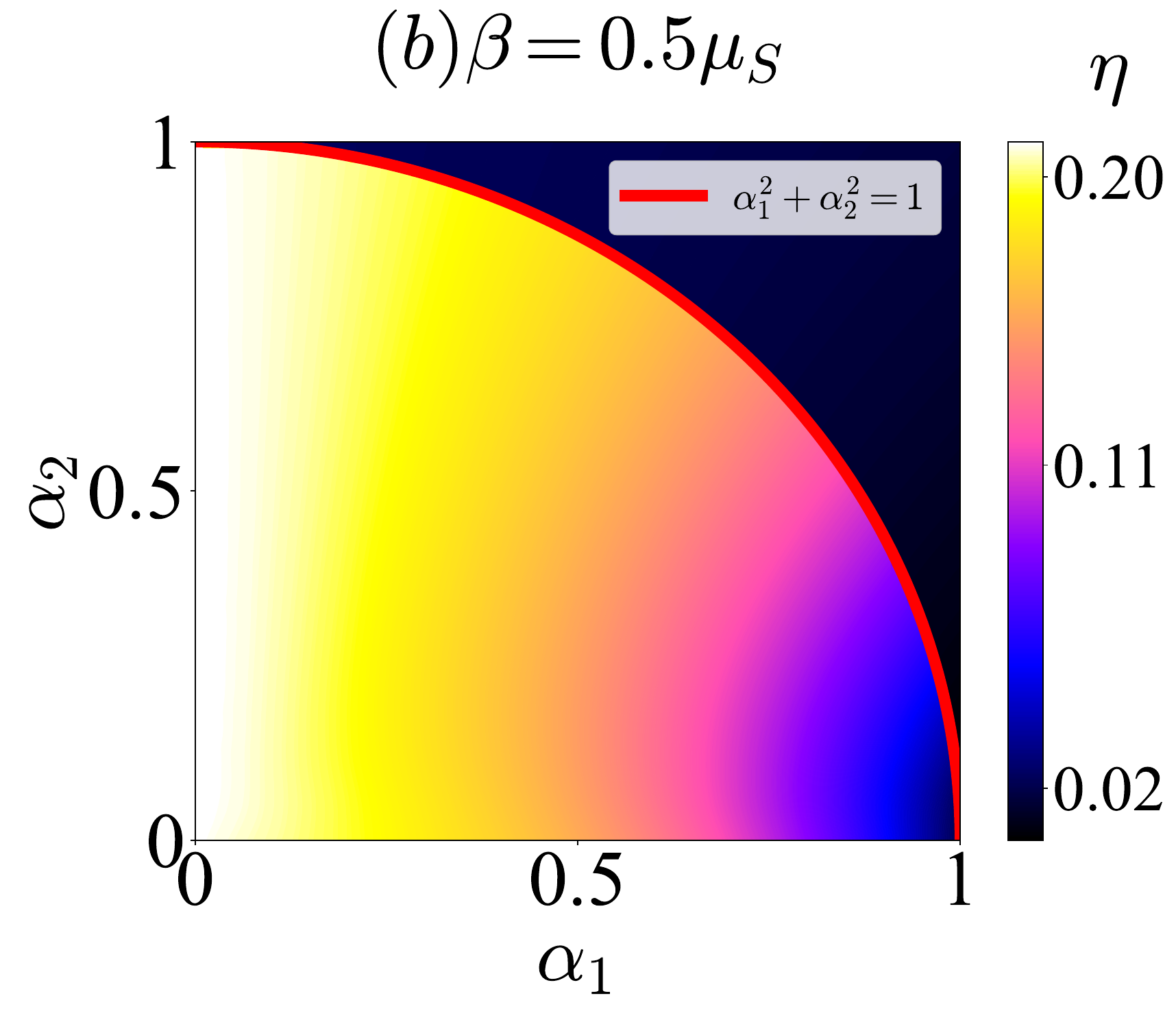}
\includegraphics[scale = 0.2]{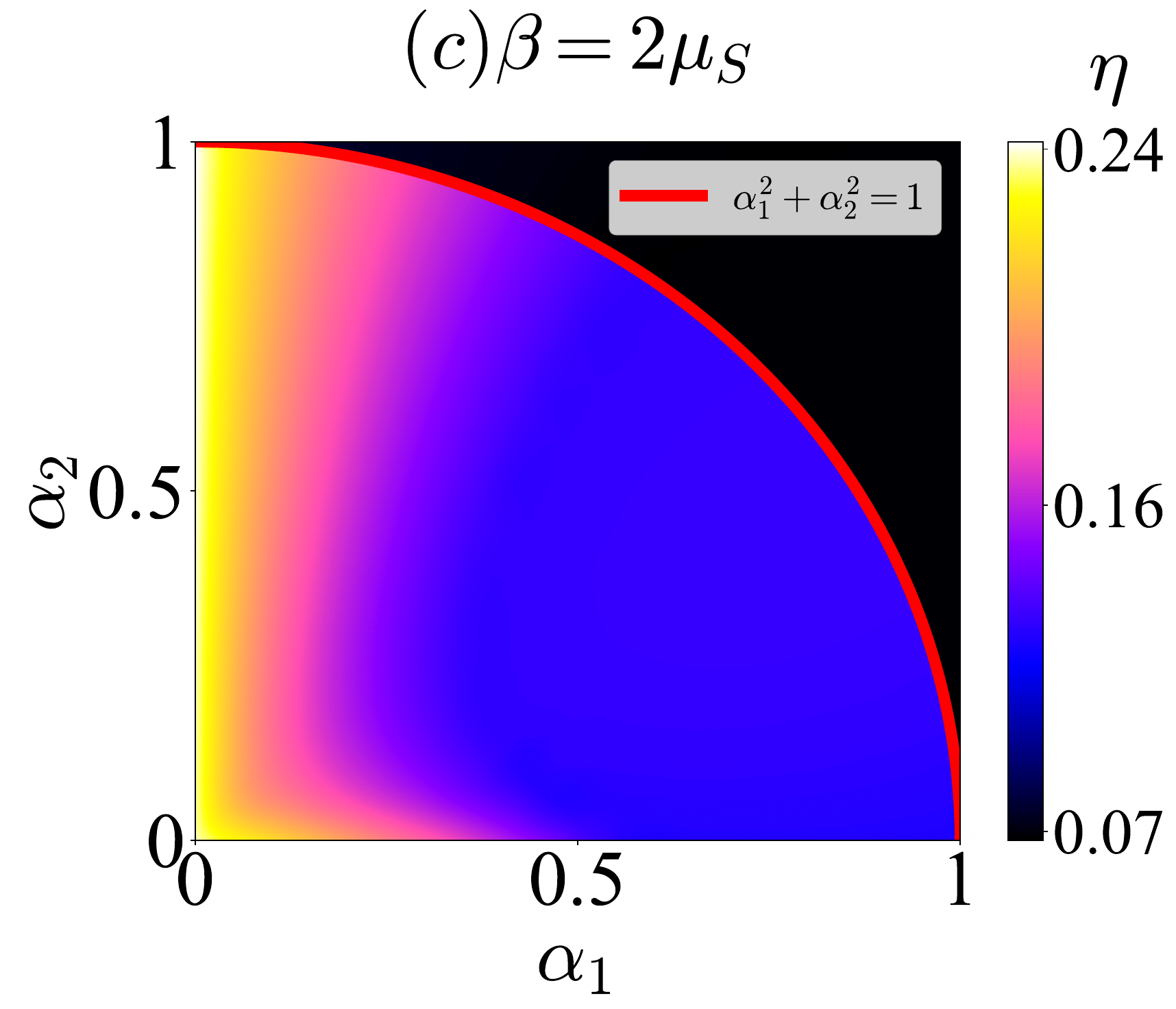}}
\caption{Density plot representing the variation of $\eta$ with $\alpha_1$ and $\alpha_2$ for (a) $\beta = 0$,  (b) $\beta = 0.5 \mu_S$ and (c) $\beta = 2\mu_S$ considering $L/L_0 = 0.1$ and $Z_0 = 0.01$. The red line corresponds to $\alpha_1^2+\alpha_2^2 = 1$, with $\alpha_c = 1$. The region below this line follows elliptic Fermi surface.}
\label{fig6}
\end{figure*} 

Although Fig. \ref{fig3} and Fig. \ref{fig4} indicate the asymmetric nature of CPR for different AM strength and orientations, a systematic investigation of the CPR asymmetry as a function of the AM layer length ($L$), barrier transparency ($Z_0$) and chemical potential ($\mu_S$) is crucial for optimizing the JDE and achieving precise control over nonreciprocal supercurrents. The dependence of $J(\phi)$ on $L$ and $\mu_S$ are shown respectively in Figs. \ref{fig5}(a), \ref{fig5}(b) and Figs. \ref{fig5}(c), \ref{fig5}(d) considering $\delta = 22.5^\circ$ and $Z_1 = 1$. For $\beta = 0.5\mu_S$, a very irregular and asymmetric CPR are noticed for both $L/L_0 = 0.1$ and $0.5$ in nearly transparent barrier configuration ($Z_0 = 0.01$). However, for an opaque barrier ($Z_0 = 0.5$), the CPR pattern are smoothen significantly yet remaining asymmetric as observed from Fig. \ref{fig5}(a).  It is due to the reason that shorter AM layer provides very weak altermagnetism, while a longer AM layer (green curves) exhibits pronounced discontinuities for both barrier transparencies, indicating strong spin filtering effect due to significantly strong altermagnetism which consequently generates higher-order harmonic contributions leading to a possible JDE. In contrast, $\beta = 2 \mu_S$ displays more conventional and nearly sinusoidal CPR pattern as seen from Fig. \ref{fig5}(a). Although an asymmetric nature of the supercurrent is observed for $Z_0 = 0.01$, but it turns out be more symmetric for $Z_0 = 0.5$ even for longer AM based Josephson junctions due to possible symmetric contributions from the Fermi lobes.  Moreover, the transition from asymmetric to symmetric CPR with increasing AM length arises from the suppression of nonlocal interference and spin-dependent phase shifts, as longer AM regions behave as diffusive scatterers, averaging out phase accumulation \cite{golubov}. Figs. \ref{fig5}(c) and \ref{fig5}(d) further analyze the role of the ratio $\mu_S/\beta$, considering three distinct cases: $\mu_S > \beta$ (single-band ISC), $\mu_S < \beta$ (double-band ISC), and $\mu_S = \beta$ (intermediate). It is noteworthy that, the supercurrent asymmetry is prominent in both the single and double band ISC cases, whereas for $\beta = \mu_S$ (blue lines), the supercurrent remains symmetric as already observed in Fig. \ref{fig3} and Fig. \ref{fig4}. This symmetry arises from the degeneracy of spin components at $\beta = \mu_S$, while for $\mu_S \neq \beta$, spin splitting occurs due to the simultaneous breaking of inversion and time-reversal symmetry \cite{cheng11}. This splitting results in an asymmetry between positive and negative supercurrents, thereby facilitating the emergence of a JDE \cite{golubov}. These results provide strong theoretical evidence for the tunable diode effect in ISC/AM/ISC junctions, offering potential applications in superconducting spintronics and dissipationless charge transport.

 \section{Josephson Diode Effect}
Fig. \ref{fig3} - Fig. \ref{fig5}, indicate that $J_c^+$ and $J_c^-$ significantly differ for strong $Z_1$, arbitrary misalignment of the Fermi surfaces in AM ($\delta$) and $\beta \neq \mu_S$. Thus it is necessary to investigate the JDE and provide diode efficiency for our proposed Josephson junction.  As already discussed due to the combined effects of altermagnetism and ISOC, the CPR deviates from the standard $J_s(\phi) = J_c \sin \phi$ relation. The presence of asymmetric terms leads to a skewed CPR:
\begin{equation}
\label{eq17}
J_s(\phi) = J_c^+ \sin(\phi + \delta) - J_c^- \sin(\phi - \delta).
\end{equation}
The Josephson diode efficiency $(\eta)$ can be defined as
\begin{equation}
\label{eq18}
\eta = \frac{|J_c^+| - |J_c^-|}{|J_c^+| + |J_c^-|}.
\end{equation}
where, the critical currents are given by
\begin{equation}
\label{eq19}
    J_c^+ = \max_{\phi} J_s (\phi), \quad J_c^- = \min_{\phi} J_s (\phi).
\end{equation}
For a normal ISC junction in the absence of AM, $J_c^+ = J_c^-$, leading to $\eta = 0$. However, in the presence of an AM with momentum-dependent spin splitting, the diode efficiency becomes finite and depends on the strengths of the physical parameters. 

\begin{figure*}[hbt]
\centerline
\centerline{
\includegraphics[scale = 0.135]{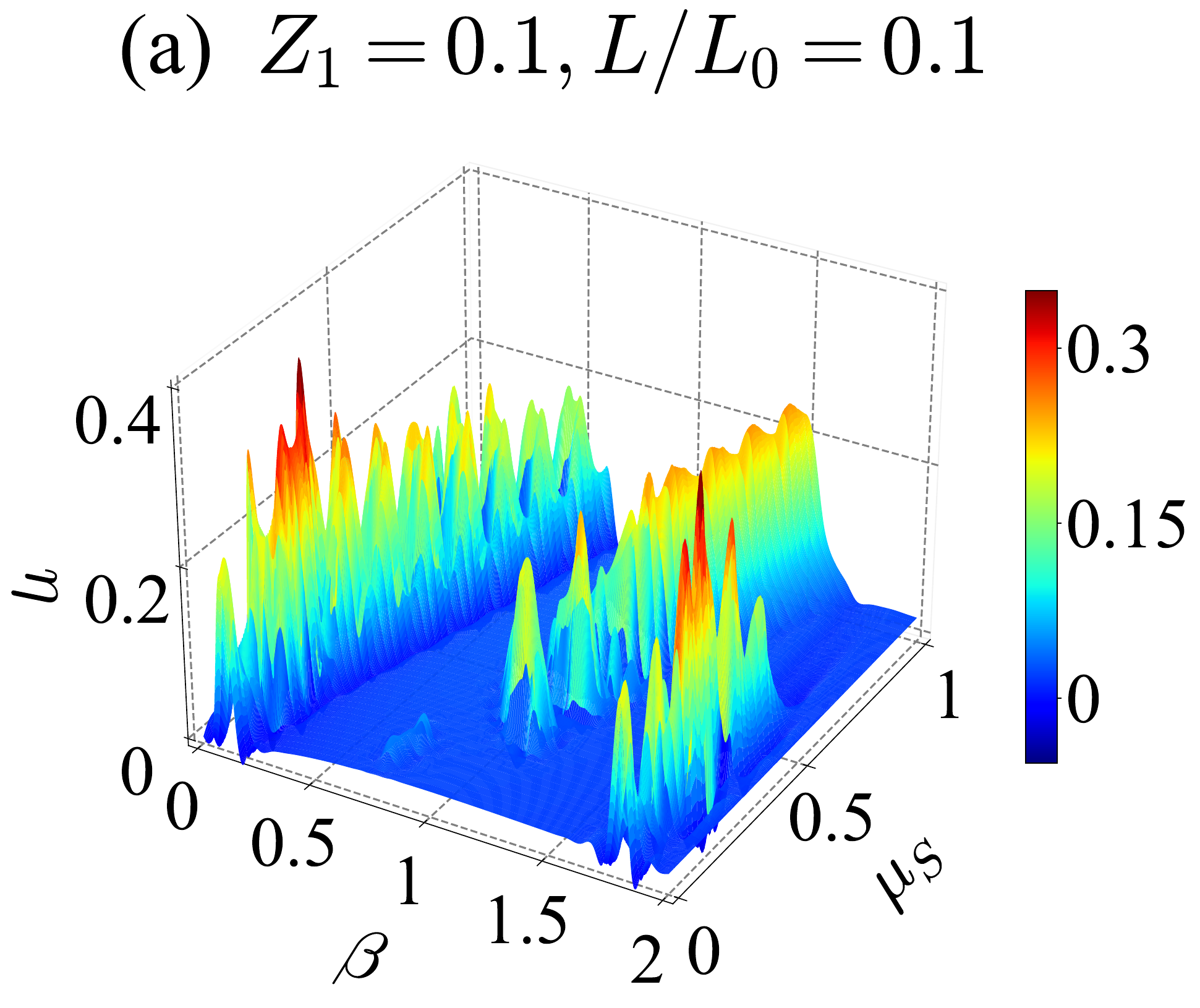}
\includegraphics[scale = 0.135]{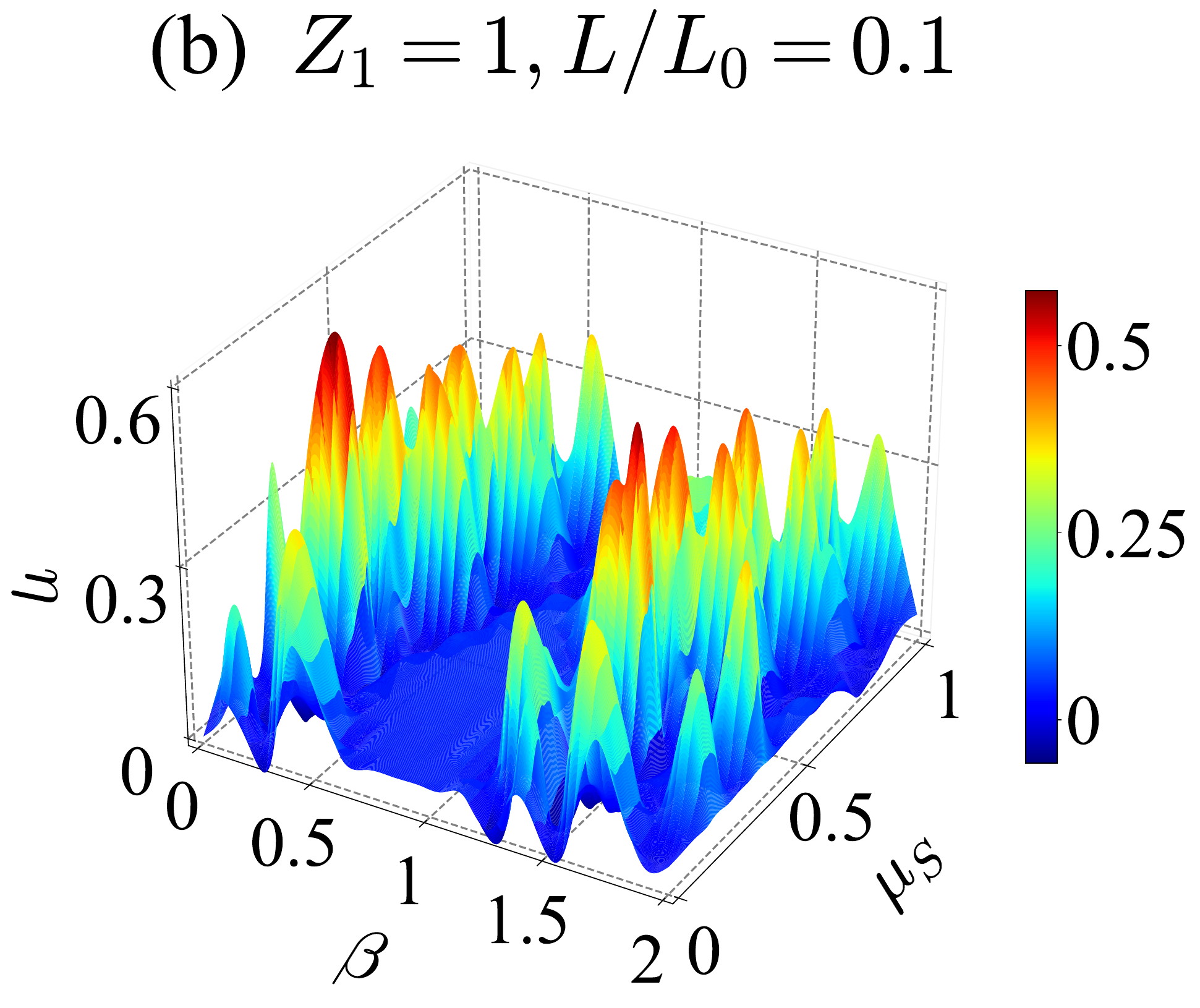}
\includegraphics[scale = 0.135]{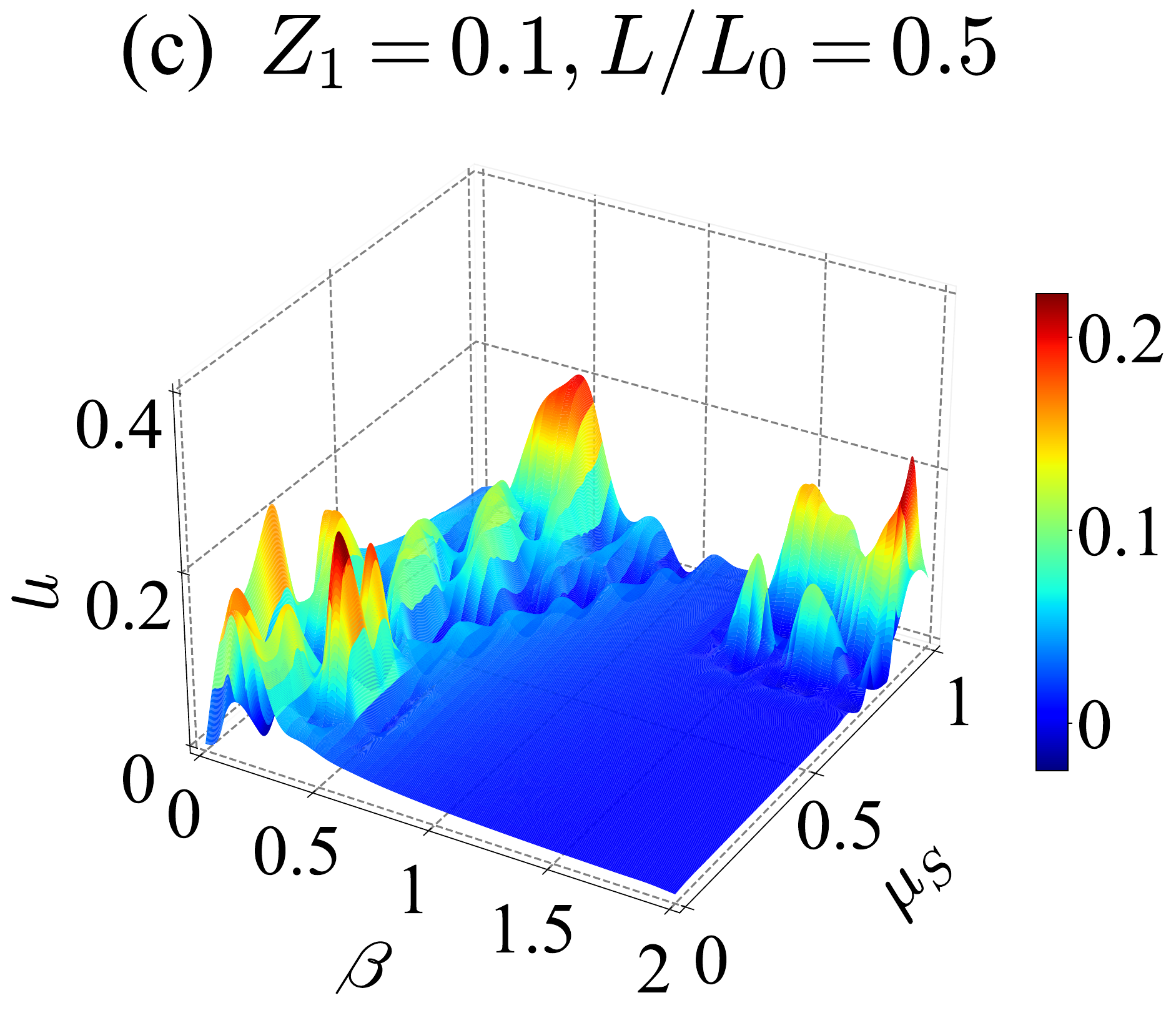}
\includegraphics[scale = 0.135]{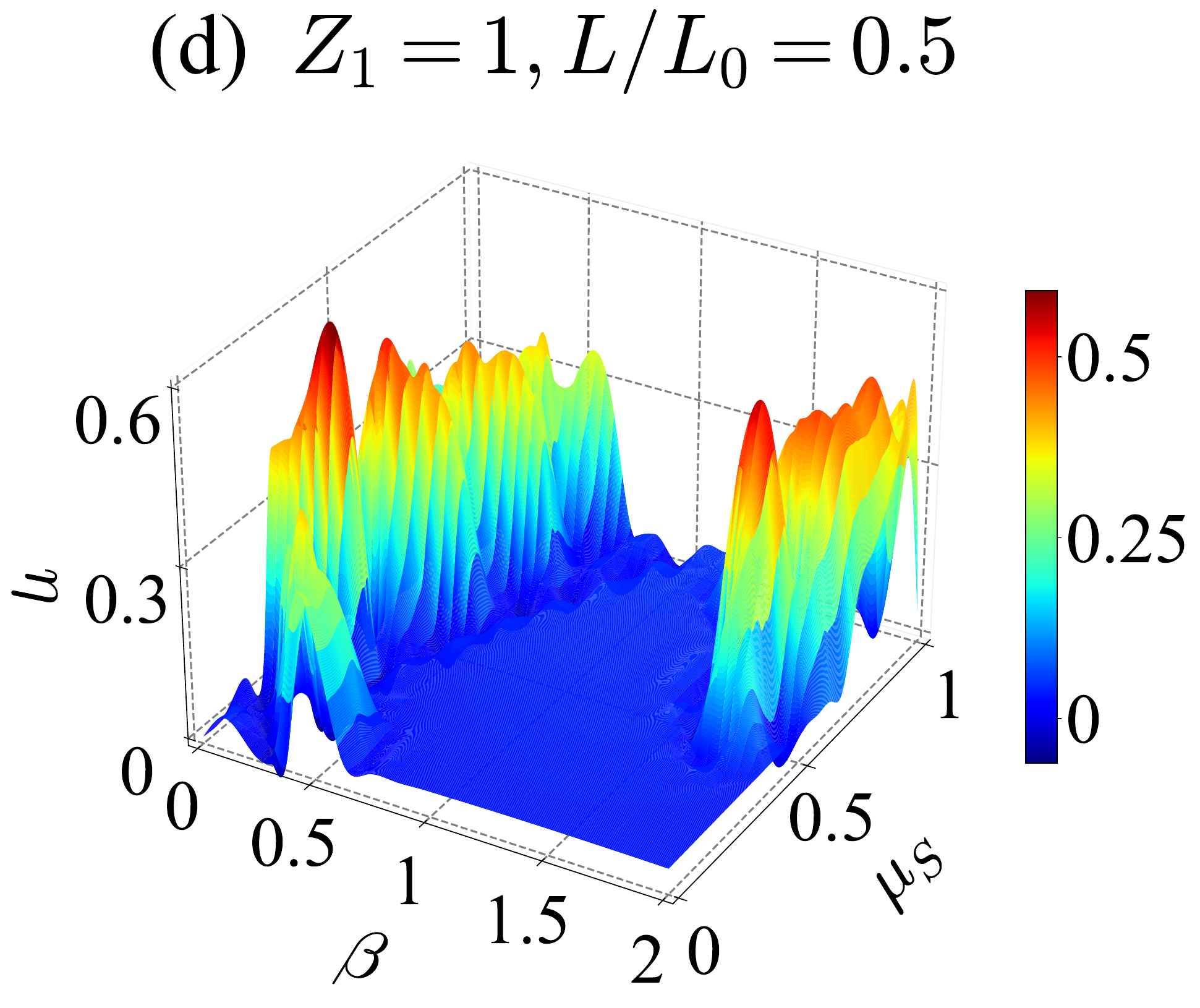}}
\caption{Variation of efficiency ($\eta$) with $\beta$ and $\mu_S$ for a junction with $Z_0 = 0.01$. The plots in (a) and (c) are for $L/L_0 = 0.1$ and $0.5$ with  $Z_1 = 0.1$ while (b) and (d) are for $L/L_0 = 0.1$ and $0.5$ with $Z_1 = 1$.}
\label{fig7}
\end{figure*} 

\subsection{Impact of the AM parameters $\alpha_1$ and $\alpha_2$ on JDE}
The parameters $\alpha_1$ and $\alpha_2$ play a crucial role in shaping the electronic properties of AM. The term $\alpha_1k_xk_y\hat{\sigma}_z$ in Eq. (\ref{eq5}) introduces a momentum-dependent spin splitting, which leads to an anisotropic band structure. Thereby it enhances non-reciprocal Andreev reflection resulting in asymmetric CPR. On the otherhand, the $\alpha_2(k_x^2 - k_y^2)\hat{\sigma}_z$ term introduces additional anisotropy by breaking the rotational symmetry, further modifying the spin-dependent dispersion.  The combined influence of these terms leads to unique nontrivial spin textures, resulting in charge and spin transport. Although we have considered different $Z_1$ values in CPR, it should be noted that $Z_1$ provides the combined strength of $\alpha_1$ and $\alpha_2$, but the interplay between them  can be understood by considering different sets of $(\alpha_1, \alpha_2)$. So, it is necessary to investigate the effect of $\alpha_1$ and $\alpha_2$ on JDE. The diode efficiency with respect to the relative values of $\alpha_1$ and $\alpha_2$ are illustrated in Fig.\ref{fig6} for a short AM length $(L/L_0 = 0.1)$. Moreover, we confine our analysis to short and moderate values of $L$ as significant asymmetry in CPR is noticed in such scenarios.  In absence of ISOC, there is a lack of broken inversion symmetry, resulting in the absence of any diode efficiency except for some particular $(\alpha_1, \alpha_2)$ combinations as seen from Fig.\ref{fig6}(a). This suggests that for $\beta = 0$, the diode effect is highly sensitive to the relative strengths of $(\alpha_1, \alpha_2)$. However, in specific regions of the $(\alpha_1, \alpha_2)$ plane, the specific $\alpha_1$ and $\alpha_2$ values provide constructive interference with ISOC, leading to resonant enhancement of the CPR asymmetry. This resonance causes a localized increase in $\eta$, forming the bright bands. 

For a single band ISC, with $\beta = 0.5 \mu_S$, the efficiency profile becomes smoother and more structured, with low $\alpha_1$ and low $\alpha_2$ values provide maximum efficiency while it reduces with the increasing values of $\alpha_1$ as seen from Fig. \ref{fig6}(b). This is because of the anisotropic band structure due to interplay of $\alpha_1$ and $\alpha_2$, resulting constructive interference thereby enhancing nonreciprocal transport for low $\alpha_1$ values while destructive interference for higher values. The suppression of efficiency suggests that moderate ISOC stabilizes the AM-ISC coupling, leading to more systematic dependence of $\eta$ on $\alpha_1$ and $\alpha_2$. Moreover, the larger region of high $\eta$ is due to high nonreciprocity in CPR as observed in Fig. \ref{fig4}, in this regime making JDE more predictable and tunable. In case of a double band ISC with $(\beta = 2\mu_S)$, the diode efficiency is more enhanced but become more localized, as seen in the shrinking of bright regions in Fig. \ref{fig6}(c) compared to Fig. \ref{fig6}(b). This suggests that while large $\beta$ enhances spin splitting and asymmetry in Cooper pair transport, it also restricts the range of $(\alpha_1, \alpha_2)$ values that support a strong diode effect. The suppression of bright regions at intermediate and high values of $\alpha_1$ implies that at strong ISOC, the system may enter a regime dominated by spin-selective superconducting pairing, which can suppress nonreciprocity outside some specific parameter regions.
The contrast between Fig. \ref{fig6}(b) and Fig. \ref{fig6}(c) highlights that while increasing $\beta$ enhances Cooper pair spin filtering, it can also reduce the range of AM configurations that support high $\eta$. This suggests that for optimal JDE, a single band ISC is most preferable.
\begin{figure}[hbt]
\centerline
\centerline{
\includegraphics[scale = 0.28]{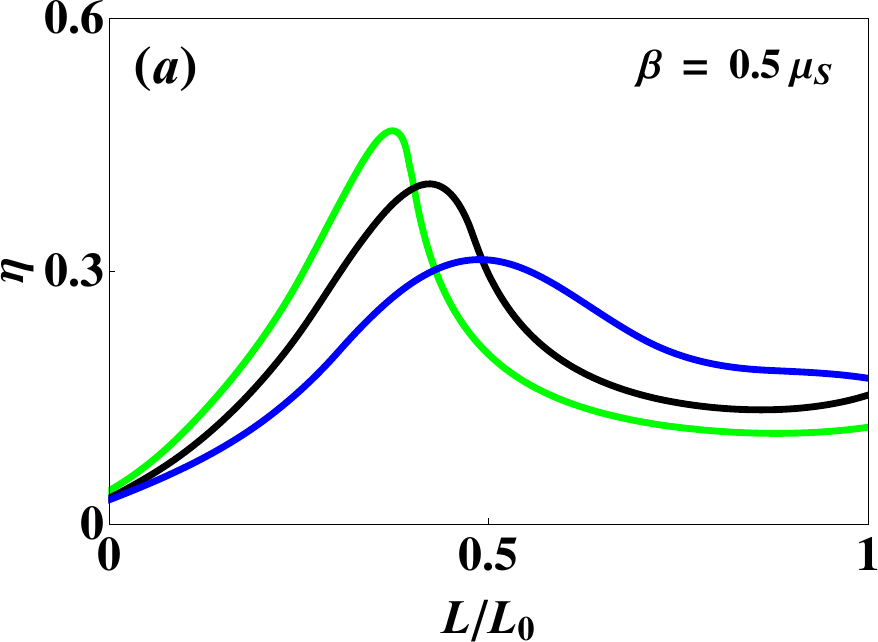}
\includegraphics[scale = 0.28]{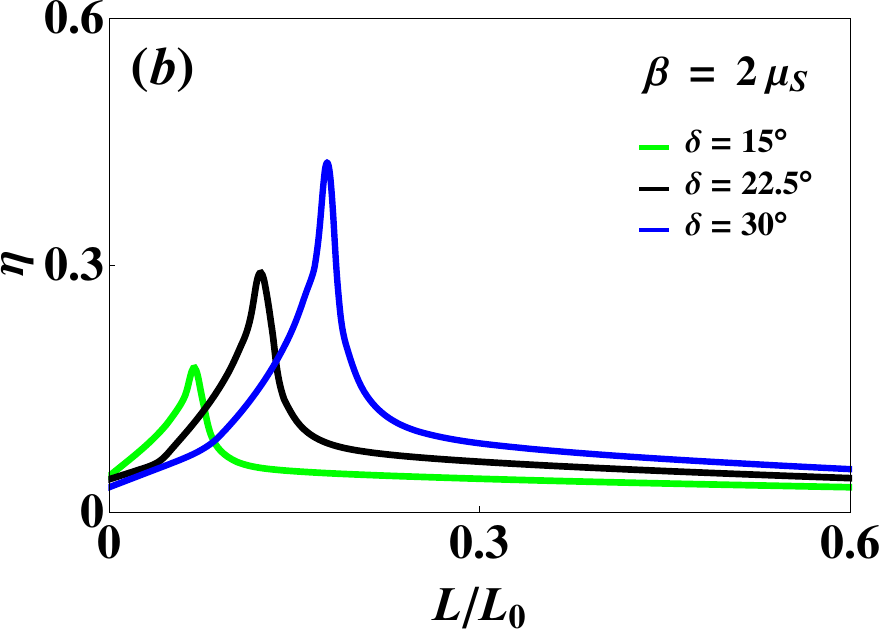} 
}
\caption{Variation of efficiency ($\eta$) with $L/L_0$ and $\mu_S$ for a junction with $L/L_0 = 0.5$. The top panel shows results for a nearly transparent barrier ($Z_0 = 0.01$) with (a) $Z_1 = 0.1$ and (b) $Z_1 = 1$. The bottom panel presents results for an opaque barrier ($Z_0 = 0.3$) with (c) $Z_1 = 0.1$ and (d) $Z_1 = 1$.}
\label{fig8}
\end{figure} 

\begin{figure*}[hbt]
\centerline
\centerline{
\includegraphics[scale = 0.55]{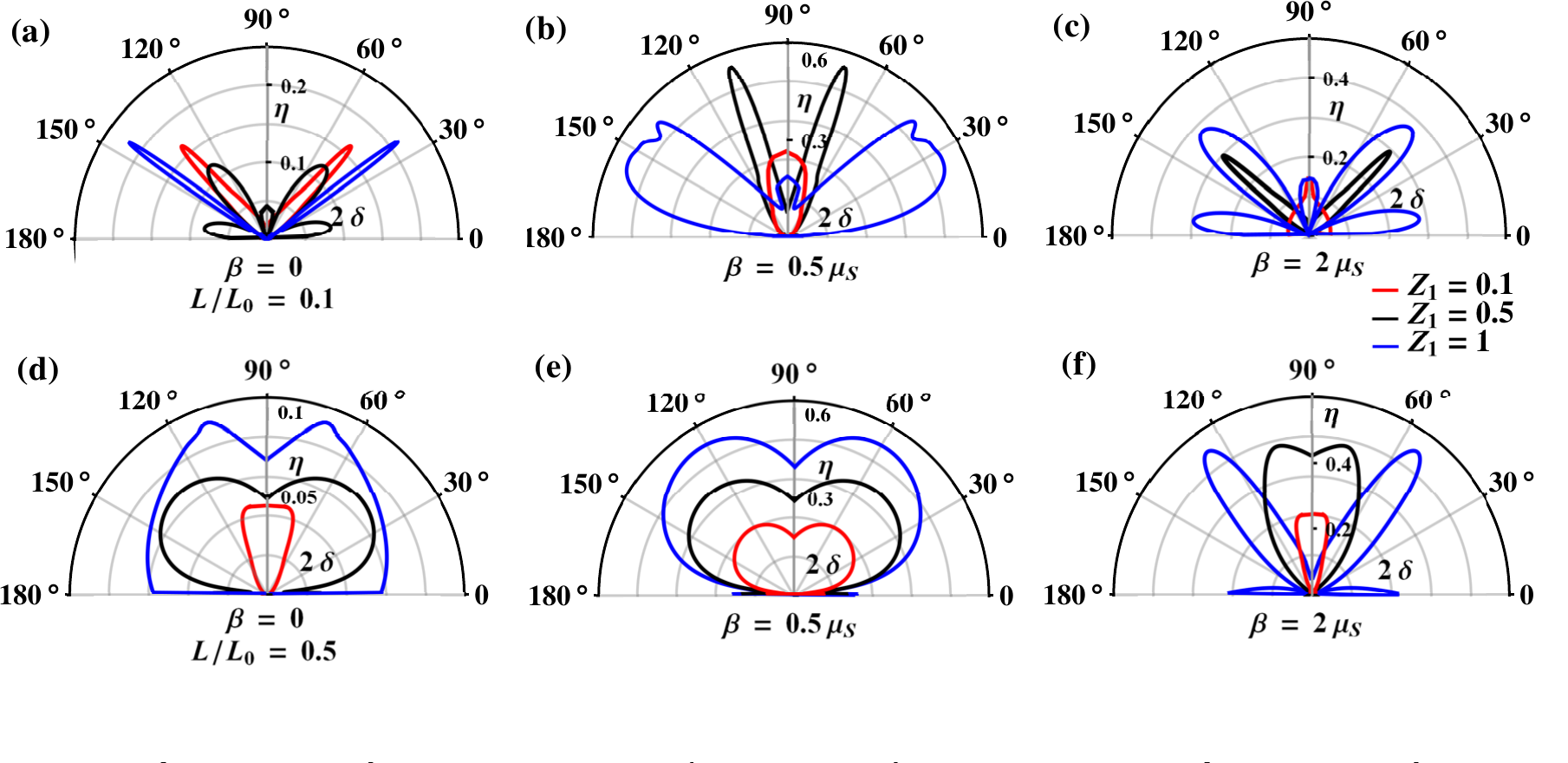}}
\hspace{0.1cm}
\vspace{-0.7cm}
\caption{Polar plot of $\eta$ with AM orientation $2\delta$ for $L/L_0 = 0.1$ (top panel) and  $L/L_0 = 0.5$ (bottom panel) considering $Z_1 = 0.1$ (Red), $Z_1 = 0.5$ (Blue) and $Z_1 = 1$ (Black) with $\mu_S = 1$ and $Z_0 = 0.01$. Plots (a) and (d) are for $\beta = 0$, (b) and (e) are for single-band ISC with $\beta = 0.5\mu_S$ while plots (c) and (f) are for a double-band ISC with $\beta = 2\mu_S$.}
\label{fig9}
\end{figure*}
\subsection{Impact of AM length, ISOC and $\mu_S$ on JDE}
Fig. \ref{fig7} presents a comparative analysis of the variation of $\eta$ as a function of  $\beta$ and $\mu_S$ for  short $(L/L_0 = 0.1)$ and moderately long $(L/L_0 = 0.5)$ junctions, under weak $(Z_1 = 0.1)$ and strong $(Z_1 = 1)$ AM strengths considering $Z_0 = 0.01$. The efficiency profile display moderate peaks for low and intermediate region of $\beta$ with considerably strong values of $\mu_S$. While for strong $\beta$ values the peaks are found for low and moderate $\mu_S$ values as observed from Fig. \ref{fig7}(a). Notably, the efficiency attains its highest values around $(\beta, \mu_S) \sim (0.2, 0.4)$ and $(1.9, 0.4)$, consistent with the trends observed in Fig. \ref{fig6}. Upon increasing $Z_1$ to $1$, the efficiency profile becomes significantly more irregular, characterized by stronger fluctuations, which suggest an enhanced sensitivity to variations in $\beta$ and $\mu_S$, as illustrated in Fig. \ref{fig7}(b). In this regime, the efficiency is significantly enhanced, reaching a maximum value of approximately $\eta \sim 0.52$ at $(\beta, \mu_S) \sim (0.21, 0.45)$ and $(1.57, 0.45)$. Furthermore, $\eta$ exhibits more pronounced peaks across for both weak and strong $\beta$ and $\mu_S$ regimes, making it a promising candidate for JDE based device. 

For moderate AM length $L/L_0 = 0.5$, the peaks in efficiency profile are significantly suppressed for $Z_1 = 0.1$, while very prominent peaks are observed for $Z_1 = 1$ as seen from Fig. \ref{fig7}(c) and Fig. \ref{fig7}(d). Notably, increasing length of the AM $L/L_0 = 0.5$ in Fig. \ref{fig7}(c) and Fig. \ref{fig7}(d) leads to pronounced efficiency modulations, particularly for a stronger AM strength $Z_1 = 1$, where efficiency maxima emerge at intermediate $\beta$ and $\mu_S$ values. These trends arise from the interplay between ABS and spin-dependent phase shifts induced by the AM, which influence the asymmetry in the CPR and, consequently, the diode efficiency.  However, in this scenario the efficiency in high $\beta$ and low $\mu_S$ regions are completely suppressed leading to nearly zero efficiency. Similar characteristics is also observed even for strong AM strength with $Z_1 =1$. So, this results suggest that ISC/AM JDE based devices with low AM length but having strong AM strength provide excellent diode efficiency. This results highlight the critical role of the ISOC and AM strength in controlling the diode effect in ISC/AM/ISC junctions.

\subsection{Impact of AM orientation on JDE}
The variation of $\eta$ with $L/L_0$  for different values of $\delta$ of the Josephson diode in an ISC/AM/ISC junction are shown in Fig. \ref{fig8}.  The left panel corresponds to an ISOC strength of $(\beta = 0.5\mu_S)$, while the right panel represent the results for $\beta = 2\mu_S$. The variation of $\eta$ with the normalized junction length $L/L_0$ demonstrates that increasing $\delta$ enhances the JDE in the strong ISOC regime. This behavior originates from the anisotropic spin splitting induced by the AM order, which modifies the spin-dependent phase accumulation of the quasiparticles propagating across the junction. For $\beta = 0.5\mu_S$, the efficiency exhibits a relatively broad peak, suggesting that the diode effect is predominantly governed by the asymmetric ABS asymmetric. However, for $(\beta = 2\mu_S)$, the efficiency profile becomes highly nonmonotonic, with a sharp peak emerging at small $L$. This enhancement in nonreciprocity can be attributed to the interplay between strong ISOC and the AM exchange field, which leads to an asymmetric current-phase relation (CPR) driven by spin-dependent interfacial phase shifts. Furthermore, the suppression of $\eta$ at larger $L$ indicates that multiple Andreev reflections contribute to the reduction of diode efficiency by averaging out the phase-sensitive effects associated with AM-induced spin filtering. The observed trends underscore the role of AM orientation in tuning the nonreciprocal Josephson response, offering a controllable mechanism for optimizing superconducting diode performance in hybrid spintronic systems. \\

To gain deeper insight into the optimal AM orientation $(\delta)$ that maximizes the diode efficiency $\eta$, we have plotted $\eta$ as a function of $\delta$ in Fig. \ref{fig9} for different values of $\beta$, specifically $\beta = 0$ (no ISOC), $\beta = 0.5\mu_S$ (single band ISC) and $\beta = 2\mu_S$ (double band ISC), , considering different values of $Z_1$ and $L$. For $\beta = 0$, a sharp efficiency peaks at specific orientations $\delta$ is observed in Fig. \ref{fig9}(a) short junctions, indicating that the AM strength significantly influences the JDE. In case of $\beta = 0.5\mu_S$, a high efficiency is observed over a broader range of $\delta$ values for $Z_1 = 1$. However, for $Z_1 = 0.5$, although the efficiency remains high, it is restricted to specific values of $\delta$. In contrast, the efficiency is nearly negligible for $Z_1 = 0.1$, as illustrated in Fig. \ref{fig9}(b).  For $\beta = 2\mu_S$, although the efficiency partially reduces but provide a number of angles at which $\eta$ is found to be maximum for $Z_1 = 1$. However, $\eta$ is still confined to specific angle for $Z_1 = 0.5$, but have minimum value for $Z_1 = 0.1$ as seen from in Fig. \ref{fig9}(c). For longer junctions $L/L_0 = 0.5$, the efficiency can be realized for a range of $\delta$ values in both $\beta = 0$ and $0.5 \mu_S$ conditions  with $\eta$ being maximum for $Z_1 = 1$ and minimum for $Z_1 = 0.1$ as observed from in Fig. \ref{fig9}(d) and Fig. \ref{fig9}(e). However, lack of ISOC leads to small $\eta$ values for all values of $Z_1$. As $\beta = 2\mu_S$. the efficiency is again confined to specific value of $\delta$ for $Z_1 = 1$ while for $Z_1 = 0.5$ JDE is observed for a wide range of $\delta$ as indicated by Fig. \ref{fig9}(f). Moreover, $\eta$ is still minimum for $Z_1 = 0.1$ in this case as well. Fig. \ref{fig9} clearly illustrates that for a better JDE precise control of AM length and proper understanding of the interplay of AM and ISOC is necessary. Also, Fig. \ref{fig9} clearly reinstates that $\eta$ is comparatively more for a single band ISC/AM than double band ISC/AM Josephson junctions. 

Our results remain consistent even in the diffusive limit. This robustness primarily arises due to two key factors: (1) The $\phi$ states and the associated anomalous supercurrent and the asymmetry in CPR are primarily originate from the symmetry of spin chirality, combined with the spin filtering and scattering effects at the AM interfaces. Since these effects are intrinsic to the system, they persist regardless of whether the transport regime is ballistic or diffusive. (2) The suppression of higher harmonics leads to a predominantly CPR for both low and high barrier transparencies, ensuring consistency across different transport regimes. In the diffusive limit, one may employ the quasi-classical Eilenberger equations \cite{miyawaki} or the Matsubara Green's function formalism \cite{pinon} to determine the Andreev bound states (ABS) and the corresponding CPR. While some qualitative differences may emerge, the overall results remain largely unaffected, exhibiting no significant deviations.

\section{Conclusions and Proposed experimental realization }

In summary, we have theoretically investigated the nonreciprocity of supercurrent in an ISC/AM/ISC Josephson junction and uncovered several anomalous phenomena, including asymmetric positive and negative supercurrents, $0 - \pi$ transitions, and also sharp CPR peaks. Our study demonstrates that a JDE can be achieved in an ISC/AM based Josephson junction in absence of an  external magnetic field. It is observed that the JDE can be easily obtained and tuned in a normal SC based Josephson junction by AM strength and orientation in absence of SOC. However, the presence of ISOC further enhances the diode efficiency. Moreover, the diode efficiency for single band ISC/AM based Josephson junction is found to be more than that of a double band ISC/AM based Josephson junction. This effect arises due to the combined effect of ISOC and broken inversion symmetry in ISC, with the broken time-reversal symmetry and rotational symmetry in the AM. Our results also suggest that the arbitrary AM orientations are most efficient for JDE while the diode effect is nearly suppressed in the vicinity of the orientations $0^\circ$ and $45^\circ$. The range of such AM orientations in case of a single band ISC is much larger than a double band ISC. Additionally, our analysis reveals that the barrier transparency and the thickness of the AM layer play very crucial roles in shaping the diode effect. An increase in barrier width suppresses higher harmonics in the supercurrent, leading to conventional sinusoidal CPR. Moreover, JDE is generally more pronounced in longer junctions compared to shorter ones. Notably, for single-band ISC/AM junction, the JDE persists regardless of the AM layer's length, whereas for a double-band ISC/AM junction, it is constrained to a specific AM thickness. 

Recent advancements in two dimensional materials, specifically transition metal dichalcogenides can act as Ising Superconductor and have been realized easily, while Altermagnetism had been observed in compounds such as RuO$_2$, MnRe, Mn$_5$Si$_3$, MnF$_2$ and La$_2$CuO$_4$. So, our proposed ISC/AM/ISC junction is well within reach of practical implementation. Moreover, phase-sensitive transport measurements and spectroscopic techniques can be employed to experimentally measure diode efficiency and asymmetric CPR in our proposed Josephson junction. Furthermore, the diode efficiency can also be determined by measuring the forward and reverse supercurrents, while SQUID interferometry can be used to probe the asymmetry in the CPR. These experimental approaches would provide direct validation of our theoretical predictions and further establish the ISC/AM/ISC junction as a promising platform for realizing high-efficiency superconducting diodes.\\
\\

\appendix
\section{Wavevectors of Arbitrary angle rotated AM}
\label{appenA}
The energy eigenvalues for an arbitrary angle rotated AM are
\begin{equation}
\label{B1}
E_{\pm} = \frac{\hbar^2 (\kappa_x^2 + \kappa_y^2)}{2m} - \mu_\text{AM} \pm \alpha_1 \kappa_x \kappa_y \pm \frac{\alpha_2}{2} (\kappa_x^2 - \kappa_y^2) 
\end{equation}
Applying $E_1 = E_2 = E_3 = E_4 = E$, the x- component of momentum for electron and hole in AM are given by 
\begin{widetext}
\begin{eqnarray}
\label{B2}
\label{B3}
\label{B4}
\label{B5}
\kappa_{e\uparrow, \pm} = \frac{-m\alpha _1 \kappa_y+\sqrt{2 m (\mu_\text{AM}+E ) \left(\hbar^2+m\alpha _2 \right) + \kappa_y^2 \{-\hbar^4+\left(\alpha _1^2+\alpha _2^2\right) m^2\}}}{(\hbar^2+m\alpha _2 )}\\
\kappa_{e\downarrow, \pm} = \frac{-m\alpha _1 \kappa_y-\sqrt{2 m (\mu_\text{AM}+E ) \left(\hbar^2+m\alpha _2 \right) + \kappa_y^2 \{-\hbar^4+\left(\alpha _1^2+\alpha _2^2\right) m^2\}}}{(\hbar^2+m\alpha _2 )}\\
\kappa_{h\uparrow, \pm} = \frac{m\alpha _1 \kappa_y+\sqrt{2 m (\mu_\text{AM}+E ) \left(\hbar^2-m\alpha _2 \right) + \kappa_y^2 \{-\hbar^4+\left(\alpha _1^2+\alpha _2^2\right) m^2\}}}{(\hbar^2-m\alpha _2 )}\\
\kappa_{h\downarrow, \pm} = \frac{m\alpha _1 \kappa_y-\sqrt{2 m (\mu_\text{AM}+E ) \left(\hbar^2-m\alpha _2 \right) + \kappa_y^2 \{-\hbar^4+\left(\alpha _1^2+\alpha _2^2\right) m^2\}}}{(\hbar^2-m\alpha _2 )}
\end{eqnarray}
\end{widetext}
where, $\alpha_1$ and $\alpha_2$ are AM strength parameters. The different AM orientations can be obtained by $\delta = \frac{1}{2}\tan^{-1}\left(\frac{\alpha_1}{\alpha_2}\right)\equiv \frac{\delta_0}{2}$. The corresponding semi-major and semi-minor axes for electron are $a = \sqrt{\frac{2m(\mu+E)}{\hbar^2-m\bar{\alpha}}}$ 
and $b = \sqrt{\frac{2m(\mu+E)}{\hbar^2+m\bar{\alpha}}}$.
where, $\bar{\alpha} \equiv \sqrt{\alpha_1^2+\alpha_2^2} < \alpha_c$ for elliptic Fermi surface.

\section{2D DOS for an arbitary angle AM}
\label{AppenB}
The general expression for the 2D density of states (DOS) is:
\begin{equation}
    N(E) = \frac{1}{4\pi^2} \int_0^{2\pi} \frac{dl}{|\nabla_\kappa E_\pm(\kappa)|}
\end{equation}
where, $dl$ is the line element along the contour for elliptic Fermi surface $\bar{\alpha}<\alpha_c$ can be written as
\begin{equation}
    dl = \sqrt{\left( \frac{d \kappa_x}{d\theta} \right)^2 + \left( \frac{d \kappa_y}{d\theta} \right)^2} d\theta 
\end{equation}
Using the parametrization,
$\kappa_x = \kappa(E, \theta) \cos \theta$,  $\kappa_y = \kappa(E, \theta) \sin \theta$, 
we have
\begin{eqnarray}
    \frac{d \kappa_x}{d\theta} = \frac{d \kappa}{d\theta} \cos \theta - \kappa \sin \theta,
\\
    \frac{d \kappa_y}{d\theta} = \frac{d \kappa}{d\theta} \sin \theta + \kappa \cos \theta.
\end{eqnarray}
Thus, the total line element becomes
\begin{equation}
    dl = \sqrt{
    \left( \frac{d \kappa}{d\theta} \cos \theta - \kappa \sin \theta \right)^2 +
    \left( \frac{d \kappa}{d\theta} \sin \theta + \kappa \cos \theta \right)^2} d\theta.
\end{equation}
The magnitude of the gradient of the energy dispersion $|\nabla_\kappa E_{\pm}|$ can be obtained as 
\begin{eqnarray}
\frac{\partial E_{\pm}}{\partial \kappa_x} = \frac{\hbar^2 \kappa_x}{m} \pm \alpha_2 \kappa_x \pm \frac{\alpha_1}{2} \kappa_y\\
\frac{\partial E_{\pm}}{\partial \kappa_y} = \frac{\hbar^2 \kappa_y}{m} \mp \alpha_2 \kappa_y \pm \frac{\alpha_1}{2} \kappa_x
\end{eqnarray}\\

Therefore, the magnitude of the gradient is
\begin{equation}
    |\nabla_\kappa E_\pm(\kappa)| = \sqrt{
    \left( \frac{\partial E_{\pm}}{\partial \kappa_x}  \right)^2 +
    \left(\frac{\partial E_{\pm}}{\partial \kappa_y}  \right)^2}.
\end{equation}
Using, $\kappa_x = \kappa_{x,e\uparrow}$ and $\kappa_y = \kappa_{y,e\uparrow}$, the DOS can be written as 
\begin{eqnarray}
    N(E) = \int_0^{2\pi} N(E,\theta) d\theta
\\
    N(E, \theta) = \frac{1}{4\pi^2} \frac{\sqrt{(d \kappa_x/d\theta)^2 + (d \kappa_y/d\theta)^2}}{|\nabla_\kappa E_\pm(\kappa)|}
\end{eqnarray}
Following the same procedure, the DOS for $e_{\downarrow}$, $h_{\uparrow}$, and $h_{\downarrow}$ can also be calculated.
\section{Matrix $\hat{\mathcal{M}}(E_\pm)$ for an ISC$|$AM$|$ISC Josephson junction}
\label{appenC}

The matrix $\hat{\mathcal{M}}(E_+)$  is determined by applying the boundary conditions given in Eqs. (\ref{eq13})–(\ref{eq16}). A comprehensive expression for the boundary matrix $\hat{\mathcal{M}}(E_+)$ in the ISC/AM/ISC Josephson junction can be written as follows:

\begin{widetext}
\setlength{\arraycolsep}{0.8pt} % Reduce column spacing
\renewcommand{\arraystretch}{1.5}
\label{A1}
\begin{equation}
\begin{small}
\hat{\mathcal{M}}(E_+)=
\left(
\begin{array}{cccccccccccccccc}
 \Delta_1 & 0 & 0 & 1 & -1 & 0 & 0 & 0 & -1 & 0 & 0 & 0 & 0 & 0 & 0 & 0 \\
 0 & \Delta _1 & -1 & 0 & 0 & -1 & 0 & 0 & 0 & -1 & 0 & 0 & 0 & 0 & 0 & 0 \\
 0 & -1 & \Delta _2 & 0 & 0 & 0 & -1 & 0 & 0 & 0 & -1 & 0 & 0 & 0 & 0 & 0 \\
 1 & 0 & 0 & \Delta _2 & 0 & 0 & 0 & -1 & 0 & 0 & 0 & -1 & 0 & 0 & 0 & 0 \\
 0 & 0 & 0 & 0 & \varphi _1 & 0 & 0 & 0 & \varphi _2 & 0 & 0 & 0 & -\Delta _7 & 0 & 0 & -\Delta _4 \\
 0 & 0 & 0 & 0 & 0 & \varphi _3 & 0 & 0 & 0 & \varphi _4 & 0 & 0 & 0 & -\Delta _9 & \Delta _6 & 0 \\
 0 & 0 & 0 & 0 & 0 & 0 & \varphi _2 & 0 & 0 & 0 & \varphi _1 & 0 & 0 & \Delta _5 & -\Delta _{10} & 0 \\
 0 & 0 & 0 & 0 & 0 & 0 & 0 & \varphi _4 & 0 & 0 & 0 & \varphi _3 & -\Delta _3 & 0 & 0 & -\Delta _8 \\
\Delta _1\mathcal{P}_1 & 0 & 0 &\mathcal{P}_2 & -i \kappa _+ & 0 & 0 & 0 & i \kappa _+ & 0 & 0 & 0 & 0 & 0 & 0 & 0 \\
 0 &\Delta _1\mathcal{P}_3  & \mathcal{P}_4 & 0 & 0 & -i \kappa _- & 0 & 0 & 0 & i \kappa _- & 0 & 0 & 0 & 0 & 0 & 0 \\
 0 &-\mathcal{P}_3 &\Delta _2\mathcal{P}_4 & 0 & 0 & 0 & i \kappa _+ & 0 & 0 & 0 & -i \kappa _+ & 0 & 0 & 0 & 0 & 0 \\
 \mathcal{P}_1 & 0 & 0 &\Delta _2\mathcal{P}_2  & 0 & 0 & 0 & i \kappa _- & 0 & 0 & 0 & -i \kappa _- & 0 & 0 & 0 & 0 \\
 0 & 0 & 0 & 0 & \varphi _1\mathcal{Q}_3  & 0 & 0 & 0 & \varphi _2\mathcal{Q}_3  & 0 & 0 & 0 & -i k_+ \Delta _7 & 0 & 0 & i k_+ \Delta _4 \\
 0 & 0 & 0 & 0 & 0 & \varphi _3\mathcal{Q}_2  & 0 & 0 & 0 & -\varphi _4\mathcal{Q}_2  & 0 & 0 & 0 & -i k_- \Delta _9 & -i k_- \Delta _6 & 0 \\
 0 & 0 & 0 & 0 & 0 & 0 & -\varphi _2\mathcal{Q}_4  & 0 & 0 & 0 & \varphi _1\mathcal{Q}_4  & 0 & 0 & i k_- \Delta _5 & i k_- \Delta _{10} & 0 \\
 0 & 0 & 0 & 0 & 0 & 0 & 0 & -\varphi _4\mathcal{Q}_1  & 0 & 0 & 0 & \varphi _3\mathcal{Q}_1  & -i k_+ \Delta _3 & 0 & 0 & i k_+ \Delta _8 \\
\end{array}
\right)
\end{small}
\end{equation}
\end{widetext}
where the parameters are define,\\
$\Delta_1=w e^{i \phi_\text{L}}$, \\
$\Delta_2=w e^{-i \phi _\text{L}}$,\\
$\Delta_3=e^{i k_+ \text{L}}$, \\
$\Delta_4=e^{-i k_+ \text{L}}$,\\
$\Delta_5=e^{i k_- \text{L}}$, \\
$\Delta_6=e^{-i k_- \text{L}}$,\\
$\Delta_7=w e^{i \left(k_+ \text{L}+\phi _\text{R}\right)}$,\\ 
$\Delta_8=w e^{-i \left(k_+ \text{L}+\phi _\text{R}\right)}$,\\
$\Delta_9=w e^{i \left(k_- \text{L}+\phi _\text{R}\right)}$, \\
$\Delta_{10}=w e^{-i \left(k_- \text{L}+\phi _\text{R}\right)}$,
\\\\
$\varphi_1=e^{i \kappa _+ \text{L}}$, \indent\indent $\varphi _2=e^{-i \kappa _+ \text{L}}$,\\
$\varphi_3=e^{i \kappa _- \text{L}}$, \indent\indent $\varphi _4=e^{-i \kappa _- \text{L}}$,\\\\
$\mathcal{Q}_1= Z_0+i(Z_1+\kappa _-)$, \indent\indent $\mathcal{Q}_2=Z_0-i(Z_1-\kappa _-)$ \\
$\mathcal{Q}_3= Z_0+i(Z_1+\kappa _+)$, \indent\indent $\mathcal{Q}_4=Z_0-i(Z_1-\kappa _+)$ \\
$\mathcal{P}_1= Z_0+i(Z_1-k_+)$, \indent\indent $\mathcal{P}_2=Z_0+i(Z_1+k_+)$ \\
$\mathcal{P}_3= Z_0-i(Z_1+k_-)$, \indent\indent $\mathcal{P}_4=-Z_0+i(Z_1-k_-)$ \\


\begin{thebibliography}{99}
\bibitem{buzdin1}
A. I. Buzdin, Rev. Mod. Phys. {\bf 77}, 935 (2005).
\bibitem{bergeret1}
F. S. Bergeret, A. F. Volkov and K. B. Efetov, Rev. Mod. Phys. {\bf 77}, 1321 (2005).
\bibitem{bergeret2}
F. S. Bergeret, M. Silaev, P. Virtanen and T. T. Heikkil\"{a}, Rev. Mod. Phys. {\bf 90}, 041001 (2018).
\bibitem{zutic}
I. Z\v{u}ti\'{c}, J. Fabian and S. D. Sarma, Rev. Mod. Phys. {\bf 76}, 323 (2004).
\bibitem{saxena}
S.S. Saxena et al., Nature {\bf 406}, 587 (2005).
\bibitem{aoki}
D. Aoki et al., Nature {\bf 413}, 613 (2001).
\bibitem{pfleiderer}
C. Pfleiderer et al., Nature {\bf 412}, 58 (2001).
\bibitem{linder101}
J. Linder and J. W. A. Robinson, Nat. Phys. {\bf 11}, 307 (2015).
\bibitem{acharjee101}
S. Acharjee and U. D. Goswami, J. Appl. Phys. {\bf 120}, 263902 (2016).
\bibitem{andreev}
A. F. Andreev, Sov. Phys. JETP {\bf 19}, 1228 (1964).
\bibitem{meng}
H. Meng, J. Wu, X. Wu, M. Ren  and Y. Ren, Sci. Rep. {\bf 6},  21308 (2016).
\bibitem{trifunovic}
L. Trifunovic, Z. Popovi\'c, and Z. Radovi\'c, Phys. Rev. B {\bf 84}, 064511 (2011).
\bibitem{annunziata}
G. Annunziata, M. Cuoco, C. Noce, A. Sudb\o, and J. Linder
Phys. Rev. B {\bf 83}, 060508(R) (2011).
\bibitem{farid}
R. Farid, D. Gazizova1, B. D. E. McNiven and J. P. F. LeBlanc, Phys. Rev. B {\bf 111}, 125118 (2025).
\bibitem{acharjee1021}
S. Acharjee, A. Boruah, N. Dutta and R. Devi,  Supercond. Sci. Technol. {\bf 36} 125014 (2023).
\bibitem{acharjee1022}
S. Acharjee and U. D. Goswami, Physica E, {\bf 135}, 114926 (2022).
\bibitem{mayer}
W. Mayer et al., Nat. Commun. {\bf 11}, 212 (2020).
\bibitem{assouline1}
A. Assouline, et. al., Nat. Commun. {\bf 10}, 126 (2019).
\bibitem{yuan1}
A. C. Yuan and S. A. Kivelson
npj Quantum Mater. {\bf 9}, 93 (2024). 
\bibitem{hirai}
T. Hirai, et al. Phys. Rev. B {\bf 67}, 174501 (2003).
\bibitem{feng1}
Z. Feng, et. al., Nat. Electron. {\bf 5}, 735 (2022).
\bibitem{occhialini}
C. A. Occhialini, et. al.,  Phys. Rev. Mater. {\bf 6},
084802 (2022).
\bibitem{betancourt}
R. D. G. Betancourt, et. al., Phys. Rev. Lett. {\bf 130}, 036702 (2023).
\bibitem{smejkal1}
L. \v Smejkal, A. B. Hellenes, R. Gonz\'alez-Hern\'ndez, J. Sinova and T. Jungwirth, Phys. Rev. X {\bf 12}, 011028
(2022).
\bibitem{smejkal2}
L. \v Smejkal, R. Gonz\'alez-Hern\'ndez, T. Jungwirth, and J. Sinova, Sci. Adv. {\bf 6}, eaaz8809 (2020).
\bibitem{yuan2}
L.-D. Yuan, Z. Wang, J.-W. Luo, E. I. Rashba, and A. Zunger,
Phys. Rev. B {\bf 102}, 014422 (2020).
\bibitem{moreno}
S. Lopez-Moreno, A. H. Romero, J. Mejia-Lopez,
and A. Muñoz, Phys. Chem. Chem. Phys. {\bf 18}, 33250
(2016).
\bibitem{cheng11}
Q. Cheng and Q.-F. Sun, Phys. Rev. B {\bf 109}, 024517
(2024).
\bibitem{sun1}
C. Sun, A. Brataas, and J. Linder, 
Phys. Rev. B {\bf 108}, 054511 (2023).
\bibitem{ouassou11}
J. A. Ouassou, A. Brataas, and J. Linder, Phys. Rev. Lett. {\bf 131}, 076003 (2023).
\bibitem{papaj}
M. Papaj, Phys. Rev. B {\bf 108}, L060508 (2023).
\bibitem{beenakker1}
C.W. J. Beenakker and T. Vakhtel, Phys. Rev. B {\bf 108}, 075425
(2023).
\bibitem{ando}
F. Ando, et. al., Nature (London) {\bf 584}, 373
(2020).
\bibitem {lin}
J.-X. Lin, et. al., Nat. Phys. {\bf 18}, 1221 (2022).
\bibitem{narita}
H. Narita, et.al, 
Nat. Nanotechnol. {\bf 17}, 823 (2022).
\bibitem{ilic}
S. Ili\'c and F. S. Bergeret, Phys.
Rev. Lett. {\bf 128}, 177001 (2022).
\bibitem{he}
J. J. He, Y. Tanaka, and N. Nagaosa, New J. Phys. {\bf 24}, 053014
(2022).
\bibitem{daido1}
A. Daido, Y. Ikeda, and Y. Yanase,  Phys. Rev. Lett. {\bf 128}, 037001 (2022).
\bibitem{daido2}
A. Daido and Y. Yanase, Phys. Rev. B {\bf 106}, 205206
(2022).
\bibitem{legg}
H. F. Legg, D. Loss, and J. Klinovaja, Phys. Rev. B {\bf 106}, 104501 (2022).
\bibitem{zinkl}
B. Zinkl, K. Hamamoto, and M. Sigrist,
Phys. Rev. Res. {\bf 4}, 033167 (2022).
\bibitem{hou}
Y. Hou, et. al., Phys. Rev. Lett. {\bf 131}, 027001
(2023).
\bibitem {picoli}
T. de Picoli, Z. Blood, Y. Lyanda-Geller, and J. I. V\"ayrynen, 
Phys. Rev. B {\bf 107}, 224518 (2023).
\bibitem{hosur}
P. Hosur and D. Palacios,  Phys. Rev. B {\bf 108}, 094513 (2023).
\bibitem{blu}
B. Lu et. al., Phys. Rev. Lett. {\bf 133}, 226002 (2024)
\bibitem{olund}
C. T. Olund and E. Zhao, Phys. Reb. B {\bf 86}, 214515 (2012)
\bibitem{nadeem}
M. Nadeem, M. S. Fuhrer  and  X. Wang, Nat. Rev. Phys. {\bf 5}, 558–577 (2023).
\bibitem{volkov1}
P. A. Volkov, et. al., Phys. Rev. B {\bf 109}, 094518 (2024).
\bibitem{zhang11}
Y. Zhang, Y. Gu, P. Li, J. Hu, and K. Jian, Phys. Rev. X {\bf 12}, 041013 (2022)
\bibitem{ilic3}
S. Ili\'c, P. Virtanen, D. Crawford, Tero T. Heikkilä1, and F. Sebasti\'an Bergeret, Phys. Rev. B {\bf 110}, L140501 (2024).
\bibitem{debnath}
D. Debnath and P. Dutta, Phys. Rev. B {\bf 109}, 174511 (2024).
\bibitem{matsuo}
S. Matsuo, et. al., Nat. Phys. {\bf 19} 1636–1641 (2023).
\bibitem{fu1}
P. H. Fu, et. al., Phys. Rev. Appl. {\bf 21}, 054057 (2024).
\bibitem{ando33}
F. Ando, et. al., Nature {\bf 584}, 373–376 (2020).
\bibitem{wei}
Y.-J. Wei, J.-J, Wang, and J. Wang, Phys. Rev. B {\bf 108}, 054521 (2023).
\bibitem{barrera}
S. C. de la Barrera, et al., Nat. Commun. {\bf 9}, 1427 (2018).
\bibitem{idzuchi3}
H. Idzuchi, et al., Nat. Commun. {\bf 12}, 5332 (2021).
\bibitem{jalouli2}
A. Jalouli, et al. J. Chem. Phys. {\bf 156}, 134704 (2022).
\bibitem{tang2}
G. Tang, et al., Phys. Rev. B {\bf 104}, L241413 (2021).
\bibitem{lu2}
J. M. Lu, et al., Science {\bf 350}, 1353 (2015).
\bibitem{saito2}
Y. Saito, et al., Nat. Phys. {\bf 12}, 144 (2016).
\bibitem{xi2}
X. Xi, et al., Nat. Phys. {\bf 12}, 139 (2016).
\bibitem{dvir2}
T. Dvir, et al., Nat. Commun. {\bf 9}, 598 (2018).
\bibitem{costanzo2}
D. Costanzo, etal., Nat. Nanotechnol. {\bf 13}, 483 (2018).
\bibitem{sohn2}
E. Sohn, et al. Nat. Mater. {\bf 17}, 504 (2018).
\bibitem{li2}
J. Li, etal., Nat. Mater. {\bf 20}, 181 (2021).
\bibitem{hamil2}
A. Hamill, et al., Nat. Phys. {\bf 17}, 949 (2021).
\bibitem{ai2}
L. Ai, et al., Nat. Commun. {\bf 12}, 6580 (2021).
\bibitem{zhu22}
Z. Y. Zhu, Y. C. Cheng, and U. Schwingenschlog, Phys. Rev. B
{\bf 84}, 153402 (2011).
\bibitem{xiao2}
D. Xiao, et al. Phys. Rev. Lett. {\bf 108}, 196802 (2012).
\bibitem{kormanyos}
A. Kormanyos, et al., Phys. Rev. B {\bf 88}, 045416 (2013).
\bibitem{cappelluti}
E. Cappelluti, et al., Phys. Rev. B {\bf 88}, 075409 (2013).
\bibitem{golubov}
A. A. Golubov, M. Y. Kupriyanov and E. Il'ichev, Rev. Mod. Phys. {\bf 76} (2004).
 \bibitem{miyawaki}
 N. Miyawaki and S. Higashitani, Phys. Rev. B {\bf 99}, 134516 (2018).
 \bibitem{pinon}
 S. Pinon, V. Kaladzhyan and C. Bena,  Phys. Rev. B {\bf 101}, 205136 (2020).
\end{thebibliography}
\end{document}